\documentclass[%
 reprint,
 amsmath,amssymb,
 aps,
]{revtex4-2}

\usepackage{graphicx}
\usepackage{dcolumn}
\usepackage{bm}
\usepackage{bbm}

\usepackage[pdftex,dvipsnames]{xcolor}
\usepackage{hyperref}
\hypersetup{frenchlinks = true,
	colorlinks,
	linkcolor={red!50!black},
	citecolor={blue!50!black},
	urlcolor={blue}
}

\usepackage{mathtools}
\usepackage{printlen}
\usepackage{enumitem}

\usepackage{siunitx}
\usepackage{stfloats}
\usepackage{braket}
\usepackage{nicematrix}

\def\mat#1{\bm{#1}}
\def\op#1{\hat{#1}}
\def\vecop#1{\vec{\op{#1}}}
\renewcommand\vec{\bm}

\def\st#1{_{\mathrm{#1}}}
\def\ut#1{^{\mathrm{#1}}}

\renewcommand{\Re}{\mathrm{Re}}
\newcommand{\fig}[0]{Fig.~}
\newcommand{\dt}[0]{{\delta t}}
\newcommand{\I}{\mathbbm 1}
\renewcommand{\O}{\mathbb O}

\newcommand{\R}{\mathcal R}

\newcommand{\W}[0]{\mathcal W}
\newcommand{\tr}[0]{\mathrm{tr}}

\newcommand{\rhotop}[0]{\op{\tilde{\rho}}}

\usepackage{ifthen}

\newcommand{\x}[1][]{%
	\ifthenelse{\equal{#1}{}}{\op x}{\op x_{#1}}%
}

\newcommand{\p}[1][]{%
	\ifthenelse{\equal{#1}{}}{\op p}{\op p_{#1}}%
}

\usepackage{xifthen}
\DeclareRobustCommand{\covar}[2]{ 
	\ifthenelse{\equal{#1}{}}{\Delta {#2} ^2}{\Delta (#1, #2)}%
}

\DeclareRobustCommand{\std}[2]{%
	\ifthenelse{\equal{#1}{}}{\Delta #2 }{\sqrt{\Delta (#1, #2)}}%
}

\renewcommand{\H}{\op H}

\usepackage{algorithm}
\usepackage{algpseudocode}

\usepackage{graphicx}
\graphicspath{ {./figs/}{./figs/fig1/}{./figs/fig2/} } 

\begin{document}

\title{Gaussian theory for estimating fluctuating perturbations \\ with back action evasive oscillator variables}

\author{Jesper Hasseriis Mohr Jensen}
\email{jhasseriis@phys.au.dk}
\address{Department of Physics and Astronomy,
Aarhus University, Ny Munkegade 120, DK-8000 Aarhus C, Denmark} 
\author{Klaus M{\o}lmer}
\email{moelmer@phys.au.dk}
\address{Aarhus Institute of Advanced Studies, Aarhus University, H{\o}egh-Guldbergs
Gade 6B, DK-8000 Aarhus C, Denmark\\
Center for Complex Quantum Systems, Department of Physics and Astronomy,
Aarhus University, Ny Munkegade 120, DK-8000 Aarhus C, Denmark}

\date{\today}

\begin{abstract}
We apply a Gaussian state formalism to track fluctuating perturbations that act on the position and momentum quadrature variables of a harmonic oscillator. Following a seminal proposal by Tsang and Caves [Phys. Rev. Lett. 105, 123601 (2010)],
Einstein-Podolsky-Rosen correlations with the quadrature variables of an ancillary harmonic oscillator are leveraged to significantly improve the estimates as relevant sensor variables can be arbitrarily squeezed while evading adverse effects from the conjugate, anti-squeezed variables. Our real-time analysis of the continuous monitoring of the system employs a hybrid quantum-classical description of the quantum probe and the unknown classical perturbations, and it provides a general formalism to establish the achievements of the sensing scheme and how they depend on different parameters.

\end{abstract}

\maketitle



%
%

\section{Introduction}
The probabilistic nature of measurements on a quantum system and the disturbance of the system by measurement back action put intriguing limits to the sensitivity of measurement schemes. For measurements aiming to resolve candidate values of a classical perturbation that influences a quantum system, measurements of the same observable at different times thus face a conundrum: too precise measurements at early times cause radical changes in the state of the quantum system, and may hence hinder or seriously disrupt later precision measurements. Pioneering works by Braginsky, Thorne, Caves, and others \cite{braginsky1980quantum, caves1980measurement, braginsky1996quantum,bocko1996on} identified so-called quantum non-demolition (QND) measurement schemes, where the same observable can be  repeatedly or continuously measured over time in a manner that accumulates sensivity and gradually projects the system on an eigenstate. For canonical position and momentum variables, this is intimately connected with the concept of squeezing, which may both be a property of the initially prepared quantum probe system and an emergent property due to the measurement process itself. Notably, in the advanced LIGO gravitational wave detection one may use squeezed input states of light to enhance the  interferometric sensing of the motion of test mass mirrors \cite{tse2019quantum}. The test mass position and momentum variables, however, must obey the Heisenberg uncertainty relation, and strong position squeezing implies strong anti-squeezing of the momentum observable and, hence, anti-squeezing of later values of the position.   

Replacing continuous probing by brief measurements every half oscillation period presents a way to persistently squeeze a definite, rotating quadrature component of the oscillator \cite{vasilakis2015generation}. Another elegant proposal employs an ancillary oscillator to form commuting pairs of observables $\hat{x}_- = (\hat{x}_1 - \hat{x}_2)/\sqrt{2}$ and  $\hat{p}_+ = (\hat{p}_1 + \hat{p}_2)/\sqrt{2}$ which may be measured with arbitrary precision and which are not coupled to their conjugate observables if the two oscillators have opposite oscillator frequencies \cite{tsang2010coherent,tsang2012evading}. The ancillary system is then referred to as a negative mass oscillator (its position changes according to the value of the negative of the momentum), and the observables $\hat{x}_-,\hat{p}_+$ are referred to as quantum free or back action free. Following \cite{tsang2010coherent,tsang2012evading}, and in independent work, use of this concept has been suggested for various scenarios \cite{woolley2013twomode,zhang2013backaction}. The commuting collective observables at the heart of the back action evasion mechanism are exactly the ones proposed in the famous foundational EPR paradox by Einstein, Podolsky and Rosen \cite{einstein1935can}, and previous proposals \cite{hammerer2009establishing} and experiments \cite{wasilewski2010quantum} employing these states are, indeed, closely related to the proposal in  Refs.~\cite{tsang2010coherent, tsang2012evading}. For a pedagogical introduction and recent experiments, see Refs.~\cite{polzik2015trajectories,moller2017quantum}.

\begin{figure}[t]
    \centering
    \newcommand{\thebottrimWigner}{-0.17}
    \newcommand{\thetoptrimWigner}{+0.20}
    \includegraphics[width=\linewidth]{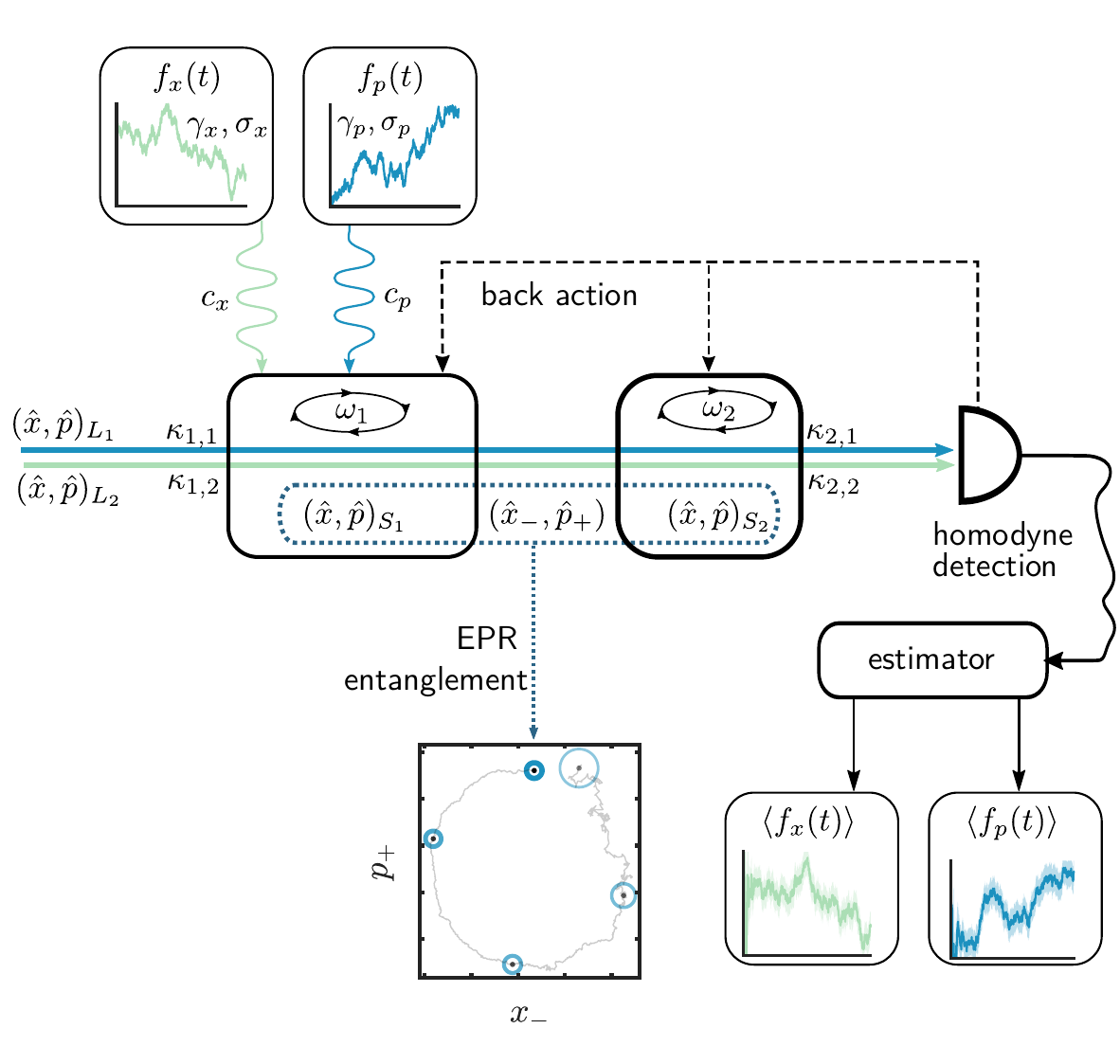} 
    \caption{
    Real time tracking of two classical perturbations $f_x(t)$ and $f_p(t)$ acting on a single oscillator $S_1$. Collective EPR variables $\x[-] \propto (\x[1]-\x[2])$ and $\p[+] \propto (\p[1] +\p[2])$  of $S_1$ and an ancillary oscillator $S_2$ are probed with two cw light beams $L_1$ and $L_2$.
    These variables commute and can be squeezed with no limits.
    Crucially, choosing opposite oscillator frequencies $\omega_1 = -\omega_2$ decouples  $\x[-]$ and $\p[+]$ from the anti-squeezed EPR variables $\p[-]$ and $\x[+]$. 
    The time evolved state has the initial oscillator ground state correlations, $\covar{}{\x[1]}=\covar{}{\p[1]} = \covar{}{\x[2]}=\covar{}{\p[2]} = 1/2$ and the gradual squeezing of the EPR observables due to the measurements benefit the estimation of the perturbations $f_x(t)$ and $f_p(t)$. The noisy data displayed in the panels come from an actual numerical simulation. 
    }
    \label{fig:setup}
\end{figure}

The use of the back action evasion mechanism is illustrated in Fig.~\ref{fig:setup}. We imagine that the  oscillator on the left is subject to perturbations that affect its position and momentum observables. By monitoring the collective EPR observables including the position and momentum of the ancillary oscillator, which is not affected by the perturbation, we squeeze both of these observables and we may infer the value of the perturbations with high precision. We sketch the probing by two light beams that undergo sequential coherent displacements proportional to the oscillator observables, and which are subsequently detected in a homodyne manner. 
As laid out in more detail in the following sections, the setup is described by the Hamiltonian
\begin{align} 
\frac{\H}{\hbar} &= \frac{1}{\hbar}(\H_{\omega_1, \omega_2} +  \H_{\vec \kappa} + \H_{c_x, c_p}) \notag \\
&\equiv \frac{\omega_1}{2} \Big(\x[S_1]^2 + \p[S_1]^2 \Big) + 
\frac{\omega_2}{2} \Big(\x[S_2]^2 + \p[S_2]^2 \Big) \notag\\
&+ 
( \kappa_{1,1} \p[S_1]  + \kappa_{2,1} \p[S_2]) \p[L_1] +
( \kappa_{1,2} \x[S_1]  + \kappa_{2,2} \x[S_2]) \x[L_2] \notag \\
&-  c_x f_x \x[S_1]  +  c_p f_p \p[S_1], \label{eq:hamiltonian}
\end{align}
where $\omega_i$ are oscillator frequencies, $\kappa_{i,j}$ are light coupling strengths, and $c_q$ are perturbation coupling strengths, respectively (for $i,j=1,2$ and $q=x,p$).
The setup is generic for a range of physical systems including mechanical and collective spin oscillators, probed by optical phase shifts or Faraday rotation angles, but we assume Gaussian states (Sec.~\ref{sec:gaussian}), including coherent, squeezed, and thermal states \cite{adesso2014continuous}, and we shall also assume that the perturbing forces or fields are governed by Gaussian statistics (Sec.~\ref{sec:estimating}).





The purpose of this article is to demonstrate the application of the Gaussian state formalism to describe both the unitary interactions and the homodyne detection for the experiment depicted in Fig.~\ref{fig:setup} (Sec.~\ref{sec:numerical}). To model the estimation of the constant or fluctuating classical perturbations, we ``quantize" their values, i.e. we associate them with QND degrees of freedom of fictitious ancillary quantum variables. This makes the Bayesian update of their classical likelihood distribution equivalent to the quantum back action on the combined quantum state, and it thus permits their inclusion in the Gaussian state formalism on equal footing with the genuine  sensor quantum observables. In particular, we characterize the entire set of quantum and classical variables by mean values and a covariance matrix, which explicitly includes estimators of the perturbations and their corresponding Gaussian variances. 

To make our presentation self contained, we present, derive and explain several elements of the physical modelling of the continuous interrogation of the oscillator systems. In this way we show how the formalism readily permits the gradual inclusion of the more specific elements of our probing scheme, and how it may be readily applied to a variety of oscillator systems and observables.
In a series of articles \cite{tsang2009time,tsang2009optimal,tsang2010optimal}, Tsang has derived equivalent equations, and we refer, in particular, to Table 1 in Ref.~\cite{tsang2009optimal}, for a summary of the connections between the theory of classical estimation theory by Kalman filters \cite{maybeck1979} and smoothers \cite{mayne1966a,fraser1969the}, and the formal elements of the quantum theory with Gaussian states and operations.     


\section{Gaussian state formalism}
\label{sec:gaussian}

Parameter estimation by continuous probing of a quantum probe system is described by the Belavkin filter  \cite{belavkin1991continuous}. This theory applies  quantum trajectory theory and generic conditional density matrices  \cite{mabuchi1996dynamical,gambetta2001state,tsang2009time}, while the restriction to Gaussian states and operations implies a significant  reduction in numerical complexity. The Gaussian description applies to the harmonic oscillator variables of the scheme presented in Fig.~\ref{fig:setup}.

\subsection{Gaussian states}
Consider a collection of $n$ canonical degrees of freedom represented by the vector of operators, 
\begin{align}
\vecop y = (\x[1],\p[1],\dots,\x[n],\p[n])^T,    
\end{align}
where  $[\x[i],\p[j]] = i \hbar \delta_{ij}$.
Any operator $\op A$ defined on this continuous variable system space, and in particular the density operator $\op A = \op\rho$ describing the quantum state, can be represented by its density operator, e.g., in the position eigenbasis $\Braket{\vec x | \op \rho | \vec x'}$, and it has an equivalent representation in terms of its Wigner function, 
\begin{align}
    \W_{\op\rho}(\vec y) = \frac{1}{(\hbar \pi)^n}\int 
    \Braket{\vec x + \vec q | \op \rho | \vec x-\vec q} e^{2i\vec p\cdot \vec q /\hbar}\, {d} \vec q,
\end{align}
where $\vec x = (x_1,\dots x_n)^T$ and $\vec p = (p_1,\dots p_n)^T$, and $d\vec q = dq_1\dots dq_n$ denotes integration over all position variable arguments $\vec q$. 
The Wigner function is a normalized quasiprobability distribution, whose value at a single point entails definite values of noncommuting variables and is hence not directly meaningful. Still the Wigner function permits calculation of expectation values as weighted ``phase space" integrals,
\begin{align}
    \braket{\op A}_{\op\rho} = \tr[\op A\op\rho] =     (\hbar\pi)^n 
    \int \W_{\op A} \W_{\op \rho}\, d\vec y, \label{eq:wigner_expectation}
\end{align}
and the special case of a projection on a definite quadrature eigenstate $\op A = \ket{y_j}\bra{y_j}$ yields the marginal density for measurements of $y_j$ by integrating all coordinates except $y_j$, 
\begin{align}
    P(y_j) = \tr[\ket{y_j}\bra{y_j} \op\rho] = \int \W_{\op\rho}\, d\vec y_{i\neq j}. 
\end{align}
The Wigner function of the remaining variables after the partial trace over the $n$'th mode, 
$\tr_{n} \rho$, is similarly given by an integral, e.g. for two modes:
\begin{align}
\mathcal{W}_{\tr_2 \op\rho}(x_1,p_1) =  \int_{} \W_{\op \rho}(x_1,p_1,x_2,p_2)\, dx_2 dp_2. \label{eq:wigner_trace}
\end{align}

The Wigner function of an arbitrary quantum state is generally a complex object, but Gaussian states, i.e. states with Gaussian Wigner functions, 
\begin{align}
    \W_{\op \rho}^{\vec m, \mat\Gamma}(\vec y) = \frac{1}{(\hbar\pi)^n}\frac{e^{-(\vec y - \vec m)^T \mat\Gamma^{-1}(\vec y - \vec m)}}{\sqrt{\mathrm{det} \mat\Gamma}}, 
    \label{eq:wigner}
\end{align}
are fully characterized by their first and second moments, i.e., by a vector of mean values and a covariance matrix,
\begin{align}
\vec m &= \braket{\vecop y}, \\
 \Gamma_{i,j} &= 
2 \Re\Big( \braket{\op y_i \op y_j} -  \braket{\op y_i} \braket{\op y_j} \Big),  \quad i,j=1,\dots,2n.
\end{align}
The variance of a quantum variable is $\covar{}{\op y_j} \equiv \covar{\op y_j}{\op y_j} = \frac{1}{2} \Gamma_{j,j}$, and we shall refer to covariance matrix elements either by their vector indices or variable names, for example 
$m_2 = m_{p_1}$ and $\Gamma_{1,4} =  \Gamma_{x_1,p_2}$.

\begin{figure}
    \centering 
    \includegraphics[width=\linewidth, trim={1cm 0cm 1cm 0cm},clip]{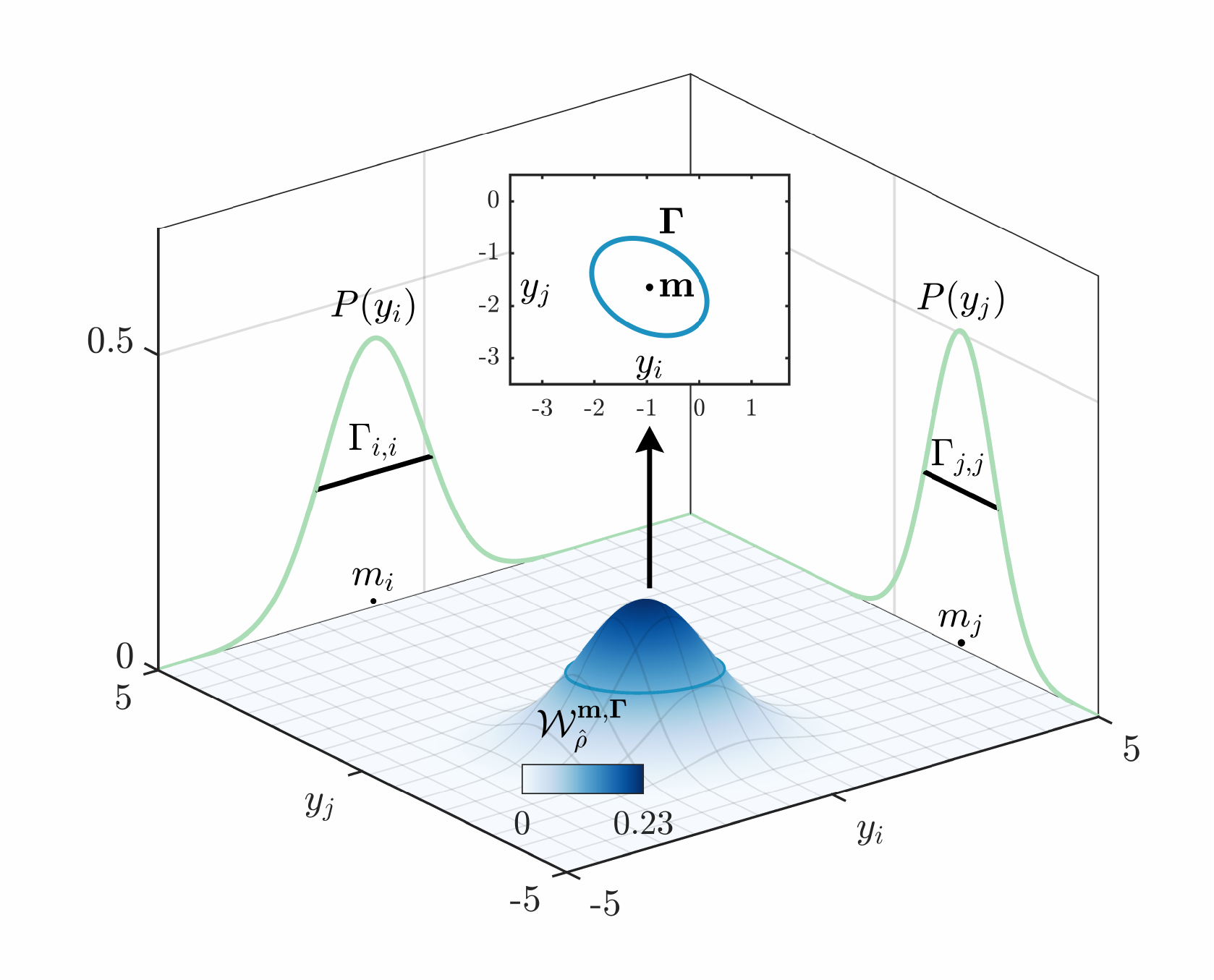}
    \caption{Illustration of a Gaussian state with a joint marginal probability density of two observables $y_i,y_j$. The density is parametrized by $\vec m = \begin{psmallmatrix} -0.95 \\ -1.65\end{psmallmatrix}$ and $\mat \Gamma = \begin{psmallmatrix}
         1.75 &  -0.43\\
         -0.43 & 1.24
    \end{psmallmatrix}$
    and the contour line at half maximum (ellipses) and mean (dots) are depicted in two dimensions in the inset. 
    The marginal densities of $y_k$ are centered on $\vec m_k$ with variances $\Gamma_{k,k}$ (for $k=i,j$).
    }
    \label{fig:WignerExample}
\end{figure}

The partial trace over some modes is effectively achieved by simply removing their corresponding entries in the covariance matrix $\mat\Gamma$ and mean vector $\vec m$, and, e.g.,  the marginal density of any single quadrature observable is a univariate Gaussian,
\begin{align}
    P(y_j) = \int \W_{\op \rho}^{\vec m, \mat\Gamma}(\vec y)\, d\vec y_{i\neq j} & 
    = \mathcal N\left(m_j, \frac{\Gamma_{j,j}}{2} \right), \label{eq:yj_marginal}
\end{align}
where $\mathcal N \propto \exp(-(y_j-m_j )^2/\Gamma_{j,j})$. 
An illustration of how marginal densities are related to joint densities is depicted for two variables  in Fig.~\ref{fig:WignerExample}.


Operations that preserve the Gaussian form of the Wigner function are fully represented by their transformation of the first and second moments. 
Knowledge of these transformations  are sufficient to describe the detection scheme in \fig\ref{fig:setup} as detailed below.

\clearpage

\subsection{Time evolution}
\label{sec:timeevolution}
Heisenbergs' equation of motion applies for all quadrature observables and for a small time step  $\dt$,
\begin{align}
\op y_j(t+\dt) \approx \op y_j(t) + \dot{\op  y}_j(t) \dt = \op y_j(t) + \frac{i}{\hbar}[\H(t), \op y_j(t)] \dt. \label{eq:heisenbergequation}
\end{align}
If the Hamiltonian $\H$ is at most quadratic in the quadrature operators $\vecop y$, as is the case in \eqref{eq:hamiltonian}, 
the unitary time evolution 
results in an affine transformation of the observables in the Heisenberg picture,
\begin{align}
\vecop y(t+\dt) = \mat S_\dt \vecop y(t) + \vec F_\dt,
\end{align}
and to first order in $\dt$ the corresponding evolution of the first and second moments yields
\begin{subequations}
\label{eqs:unitary}
\begin{align}
    \vec m(t+\dt) &= \mat S_\dt \vec m(t) + \vec F_\dt, \label{eq:unitary_mean}\\
    \mat \Gamma(t+\dt) &= \mat S_\dt \mat \Gamma(t) \mat S_\dt^T. \label{eq:unitary_covariance}
\end{align}
\end{subequations}
These update equations can be readily generalized to incorporate dissipation and noise by appropriately modifying $\mat S_{\dt} \rightarrow \mat D_{\dt}\mat S_{\dt}$, and adding extra diffusion terms, in the form of a matrix $\mat L_{\dt}$, in the update of $\mat \Gamma$, respectively. See Sec.~\ref{sec:physical} for the specific application to the setup in Fig.~\ref{fig:setup}.

\subsection{Measurement of some quadratures observables}
Suppose the system is divided into subsystems A and B with $n_A$ and $n_B$ oscillator modes, such that we may write
$\vecop y = \begin{psmallmatrix} \vecop y_A \\ \vecop y_B \end{psmallmatrix}$ where $\vecop y_A = (\x[1], \p[1], \dots, \x[n_A], \p[n_A])^T$ is of length $2n_A$
and $\vecop y_B = (\x[n_A+1], \p[n_A+1], \dots, \x[n], \p[n])^T$ is of length $2n_B$.
The Gaussian moments may then be similarly divided
\begin{align}
    \vec m &= \begin{pmatrix}\vec a \\ \vec b \end{pmatrix}, \qquad
    \vec \Gamma = \begin{pmatrix}
                \mat A & \mat C \\
                \mat C^T & \mat B
    \end{pmatrix}. \label{eqs:subsystemAB}
\end{align}
$\mat A$ is the $2n_A \times 2n_A$ covariance matrix for the variables $\vecop y_A$, 
$\mat B$ is the $2n_B \times 2n_B$ covariance matrix for the variables $\vecop y_B$, 
while $\mat C$ is the $2n_A \times 2 n_B$ covariance matrix describing the correlations, or entanglement, between the two subsystems A and B. 

We already discussed how the reduced density matrix and Wigner function of one subsystem is obtained by retaining only the relevant mean values and covariance elements, say, $\vec a$ and $\mat A$, when B is traced out. Consider now instead the state of the same subsystem, but conditioned on a projective quadrature measurement of all $n_B$ oscillator modes in subsystem B. If the two sets of subsystems are correlated, i.e. $\mat C$ is nonzero, the outcome of this measurement  will influence the resulting state of subsystem A through measurement back action. This can be understood from the inset in Fig.~\ref{fig:WignerExample} where a measurement of the variable $y_j$ causes the $y_i$ distribution to have equal values at the intersection of the contour ellipse and a horizontal line at the random outcome $y_j=y_j\ut{meas}$. 

Assuming classical (commuting) variables, for the measurement of  $\vec y_B$, the restriction of the corresponding arguments in \eqref{eq:wigner} to their measured values is a Gaussian function of the remaining variables $\vec y_A$ with the coefficients of the appropriate block submatrix of $\mat \Gamma^{-1}$. The corresponding reduced covariance matrix, in turn, is the inverse of that submatrix. By linear algebra, the inversion of the block matrix $\mat \Gamma$ \eqref{eqs:subsystemAB}, thus yields the conditional covariance matrix, $\mat A\ut{cond} = \mat A - \mat C \mat B^{-1} \mat C^T $ and vector of mean value,  $\vec a\ut{cond} = \vec a +\mat C \mat B^{-1} (\vec y\ut{meas}-\vec b)$ \cite{eaton1983multivariate}.

In the quantum setting, we cannot simultaneously measure both but only one of the $x$ or $p$ quadratures of the
$\vecop y_B$ variables while the canonically conjugate, unmeasured  quadrature variables become completely uncertain due to the Heisenberg uncertainty relation. In this case we have recourse to an appropriately modified transformation of the conditional
moments for subsystem A,
\begin{subequations}
\label{eqs:measurement}
\begin{align}
    \mat A\ut{cond}   &= \mat A - \mat C(\mat \Pi \mat B \mat \Pi)^- \mat C^T,  \label{eq:Ameasure}\\
    \vec a\ut{cond} &= \vec a + \mat C (\mat \Pi \mat B \mat \Pi)^- \vec \Delta\ut{meas}_B,  
\end{align}
where $(\cdot )^-$ denotes the Moore-Penrose pseudoinverse of its matrix argument (see \cite{eisert2002distilling} for a general discussion of the back action of quadrature measurements on Gaussian states).
The projection matrix $\mat \Pi$ is a block matrix formed by subprojectors 
\begin{align}
    \mat \pi_{i} \in \left\{ \mat \pi_x, \mat \pi_p \right\} \equiv 
    \left\{ \begin{pmatrix} 1 & 0 \\ 0 & 0   \end{pmatrix}, \begin{pmatrix} 0 & 0 \\ 0 & 1   \end{pmatrix} \right\}
\end{align}
assuming the measured quadrature for each mode $i=1, \dots, n_B$ in $\vecop y_B$ is either the position or momentum.
The measurement outcomes $q_i\ut{meas}=(q_1, \dots, q_{n_B})$ form a vector that can be written as the diagonal elements of a block matrix $\mathrm{diag}\left(\oplus_{i} q_i\ut{meas}  \mat \pi_{i}\right)$, where $q \in \{x,p\}$ is the quadrature designated by $\mat \pi_i$. Defining a similar vector of the expectation values $\Braket{\op q_i}$ of the same quantities (given by the corresponding entries of $\vec b$), we define their difference  \begin{align}
    \vec \Delta\ut{meas}_B = \mathrm{diag}\left(\oplus_{i} (q_i\ut{meas} - \Braket{\op q_i}) \mat \pi_{i}\right). 
\end{align}
See Sec.~\ref{sec:physical} for the application to the setup in Fig.~\ref{fig:setup}.
\end{subequations}


\subsection{Continuous monitoring}
\label{sec:continuousmonitor}
We are interested in the case where the subsystem B, used to probe subsystem A,
forms a continuous wave light beam in a coherent state, consistent with a Gaussian state description. To encompass both the continuous unitary evolution and evolution due to the interaction and the measurement back action we represent a continuous probe beam as a ``train" of very short segments of duration $\dt$, each treated as a single mode, and interacting sequentially with system A \cite{madsen2005continuous}.
Each field segment is initially in the same (trivial) coherent field state and is not correlated with system A prior to their interaction. The assumption of coherent states implies that there are no correlations between the incident light segments.  

Once a light segment arrives, it takes the role as subsystem B in Eqs.~\eqref{eqs:subsystemAB}  with covariance matrices and mean values,
\begin{subequations}
\label{eqs:segment}
\begin{align}
    \mat B &\rightarrow \I_{2 n_B\times 2n_B}, \\
    \mat C &\rightarrow \O_{2n_A\times 2n_B}, \\
    \vec b & \rightarrow \O_{2n_B \times 1},
\end{align}
where $\I$ and $\O$ are the identity and zero matrices of the indicated dimensions. The system and the light segment interact briefly 
by Eqs.~\eqref{eqs:unitary}, and the latter is subsequently subject to a homodyne measurement with an associated back action on system A given by Eqs.~\eqref{eqs:measurement} \footnote{The pseudoinverse of $B$, is approximated
to lowest order by $(\mat \Pi \mat B \mat \Pi)^- \approx \mat \Pi$, with corrections of order $O(\dt)$ due to the interaction.}
whence the projected field quadrature eigenstate factors and disappears from the description.
The effective $t\rightarrow t+ \dt$ evolution of system A can hence be summarized 
\begin{alignat}{2}
\vec a(t) &\overset{\eqref{eqs:unitary}}{\rightarrow} \vec a(t+\dt) &&
\overset{\eqref{eqs:measurement}}{\rightarrow}
\vec a\ut{cond}(t+\dt) \equiv \vec a(t+\dt),  \\
\mat A(t) &\,\rightarrow \mat A(t+\dt) &&\,\rightarrow \mat A\ut{cond}(t+\dt) \equiv \mat A(t+\dt). 
\end{alignat}
\end{subequations} 
This process is then repeated with the next ``fresh" incident light segments, and application of Eqs.~\eqref{eqs:segment}
yields the continued evolution of subsystem A due to the interactions and accumulation of probe measurement data. 


The continuous detection record that conditions the dynamics,
\begin{subequations}
\label{eq:detectionrecord}
\begin{align}
\vec q_B\ut{meas}(t)=(q_1, \dots, q_{n_B}),    
\end{align}
may be extracted from an actual experiment or, 
in the event of a purely numerical simulation, be sampled from a normal distribution
\begin{align}
 q_i \sim \mathcal N(\Braket{\op q_i}, \covar{}{\op q_i}) \approx \mathcal N(\braket{\op q_i}, 1/2).   
\end{align}
The expectation value $\braket{\op q_i}$ and variance $\covar{}{\op q_i}$ are elements of $\vec b$ and $\mat B$ after the interaction, respectively, and the approximation, $\covar{}{\op q_i} = 1/2$ holds for the infinitesimal interaction with light segments of short duration.
\end{subequations}
\\


We note that the covariance matrix update \eqref{eq:Ameasure} does not depend on the actual measurement outcome, and in the limit of infinitesimally small time steps, 
$\mat A$ it becomes the solution of a Ricatti matrix differential equation \cite{madsen2005continuous}, 
\begin{align}
    \dot{\mat A} &= \lim_{\dt \rightarrow 0^+} \frac{\mat A\ut{}(t+\dt) - \mat A(t)}{\dt} \\
    &\equiv \mat G - \mat D \mat A - \mat A\mat E - \mat A\mat F\mat A,
\end{align}
where the matrices $\mat G, \mat D, \mat E, \mat F$ are determined from the physical interactions leading to the dynamics. 
The nonlinear matrix Ricatti equation can be decomposed into $\mat A = \mat W \mat U^{-1}$ where $\mat W$ and $\mat U$ are solutions to two linear matrix equations $\dot{\mat W} = -\mat D\mat W + \mat G\mat U$ and $\dot {\mat U} = \mat F \mat W + \mat E\mat U$. 
Expressions for the matrices $\mat G, \mat D, \mat E, \mat F$ for our specific system are presented in the Supplementary Material \cite{supplementary}. 




\medskip


\section{Estimation of classical perturbations with continuous monitoring}
\label{sec:estimating}
A perturbation  of the oscillator system, caused e.g., by classical forces or external fields $f_i(t)$ will result in a proportional displacement of the oscillator quadratures. In this article, we assume that each of the, say, $n_f$ perturbations are individually characterized by a time-dependent Ornstein-Uhlenbeck (OU) process, 
\begin{align}
    df_i(t) = -\gamma_i f_i(t) dt + \sqrt{\sigma_i} dW_i(t), \quad i=1,\dots, n_f. \label{eq:OU} 
\end{align}
The OU process is damped with a rate $\gamma_i$ and undergoes diffusion governed by a diffusion constant $\sigma_i$ and stochastic Wiener increments, $dW_i(t)\sim \mathcal N(0,dt)$. 
We may readily include the effect on the quantum observables of such a \textit{known} time dependent perturbation through the appropriate entries in $\vec F_\dt$ in Eq.~\eqref{eq:unitary_mean}, and Section~\ref{sec:continuousmonitor} provides the quantum dynamics \eqref{eqs:segment} conditioned on any specified perturbation and an observed  detection record \eqref{eq:detectionrecord}.

However, we wish to estimate an \textit{unknown} time dependent perturbation from the measurements.
Hence rather than the $f_i$'s 
being known we represent them by probability distributions 
and infer their evolution due to the acquisition of measurement data. This is done conveniently by   incorporating the evolution of these distributions into the already established quantum formalism.

Equation~\eqref{eq:OU} is equivalent with a Fokker-Planck equation 
describing the probability density of the value of $f_i(t)$, which takes a  Gaussian form. We shall use this fact to enable a formal description of the unknown perturbations at the level of Gaussian Wigner functions and quantum density operators.

\subsection{Filtering}


\subsubsection{Quantum-classical hybrid formalism}
It is convenient to introduce for each perturbation $f_i$ an ancillary quantum operator $\op f_i$ and specify its eigenstates, $\op f_i \ket{f_i} = f_i \ket{f_i}$, with eigenvalues for each possible value of $f_i$. 
In this way, a classical probability density $P({\vec f}=(f_1,f_2, ... f_{n_f}))$ can be represented as a quantum state 
in an incoherent mixture of eigenstates, $\op \chi = \int d \vec f  \ket{\vec f}\bra{\vec f} P({\vec f})$ with $\ket{\vec f} = \otimes_{i = 1}^{n_f}  \ket{f_i}$.

We may then define an augmented density operator [note the ``$\sim$"] on the joint space of ancillary and genuine quantum variables, 
\begin{align} 
  \op {\tilde \rho} = \int d\vec f \, \ket{\vec f}\bra{\vec f} \otimes \op \rho_{\vec f}, \label{eq:rhotilde}
\end{align}
where $\rhotop(0) = \op \chi \otimes \op \rho(0)$ for some initial state $\op \rho(0)$ on the space of genuine quantum variables
\footnote{In the continuous variable position representation,
\begin{align*}
\op \rho_{\vec f} = \int d\vec x d\vec x'  \, \rho_{\vec f}^{\vec x \vec x'} \ket{\vec x  } \bra{ \vec x' }. 
\end{align*}
}.  
In Eq.~\eqref{eq:rhotilde} the probability density $P(\vec f)$ is absorbed in the norm of $\op \rho_{\vec f}$ and can be retrieved by standard quantum expressions, 
\begin{align}\label{trtr1}
    P(\vec f) &= \tr[\ket{\vec f}\bra{\vec f} {{\rhotop}}_{}] = \tr[\op \rho_{\vec f}],
    \end{align}
    \begin{align}\label{trtr2}
    P(f_j) &= \tr[\ket{f_j}\bra{f_j} {{\rhotop}}_{}] = \int d \vec f_{i\neq j}\, \tr[\op \rho_{\vec f}],
\end{align}
where $\vec f_{i\neq j}$ denotes all components of $\vec f$ except $f_j$, and ``tr" denotes the trace over all degrees of freedom of its operator argument which differ in the second and last terms of Eqs.~\eqref{trtr1} and \eqref{trtr2}. 
The time evolution of each $\op \rho_{\vec f}$ is conditioned on the value of $\vec f$ and on the measurement record $\vec q$ and, hence, the classical and quantum degrees of freedom become correlated. 
While the full state $\rhotop$ is renormalized after each measurement, the measurement back action leads to a formal redistribution of norm among the individual $\op \rho_{\vec f}$'s and hence the classical probability densities. This occurs in a manner fully equivalent with Bayes' rule \cite{supplementary}, so that outcomes that occur with higher (lower) probability for given values of $\vec f$, cause an increase (decrease) in the corresponding state components and hence the inferred likelihood for these values.

\subsubsection{Gaussian states}
While the augmented quantum state description of classical and quantum variables can in principle be employed with general interactions \cite{mabuchi1996dynamical,gambetta2001state,normann2021quantum}, the Gaussian description of both the quantum systems and the unknown classical perturbations, permits an almost straightforward application of the mean value and covariance matrix formalism to the estimation of ${\vec f}$. 


To admit the unknown classical perturbations into the quantum Wigner function description
we introduce an effective Gaussian Wigner function in the form of Eq.~\eqref{eq:wigner}, $\mathcal W_{\op{\tilde{\rho}}}^{\vec {\tilde m}, \mat {\tilde{\Gamma}}}(\vec { \tilde {y}})$
with $\vec { \tilde {y}} = \begin{psmallmatrix} \vec {{y}} \\ \vec{f} \end{psmallmatrix}$
(For a discussion of how this function is consistent with the full Wigner function representation see Ref.~\cite{supplementary}.) 
The formal probability densities 
\begin{align}
    P(f_j) = \int d\vec x d\vec p d\vec f_{i\neq j}\, \mathcal{W}_{\op{\tilde {\rho}}}^{\vec {\tilde m}, \mat {\tilde{\Gamma}}}(\vec {{y}},\vec{f}), \label{eq:probability_wigner_filtering}
\end{align}
are fully characterized by the corresponding elements in the mean vector $\vec{\tilde {m}}$ and covariance matrix $\mat{\tilde {\Gamma}}$ c.f. Eq.~\eqref{eq:yj_marginal}.  

The first and second moments of the ancillary and the genuine quantum observables are in each time step first propagated similarly to Eqs.~\eqref{eqs:unitary}:
\begin{subequations}
\begin{align}
    \vec {\tilde m}(t+\dt) &= \mat {\tilde{D}}_{\dt} \mat {\tilde S}_\dt \vec {\tilde m}(t), \\
    \mat {\tilde\Gamma}(t+\dt) &= \mat {\tilde{D}}_{\dt} \mat {\tilde S}_\dt \mat {\tilde \Gamma}(t) \mat {\tilde S}_\dt^T\mat {\tilde{D}}_{\dt}^T + \mat {\tilde L}_{\dt},
\end{align}
\end{subequations}
where the matrix $\mat{\tilde S}_{\dt}$ incorporates the displacement of the oscillator observables due to the unknown  perturbations. The friction and diffusion terms in the time dependent OU processes Eq.~\eqref{eq:OU} can be explicitly incorporated in our probabilistic treatment by average damping terms in the matrix $\mat {\tilde{D}}_{\dt}$ acting on the mean values and covariances and diffusion represented by the inhomogeneous term $\mat {\tilde L}_{\dt}$ in the equation for the covariance matrix.
See Sec.~\ref{sec:physical} for application to the setup in Fig.~\ref{fig:setup}.

The homodyne measurements on the field probe eliminate the field quantum variables $\vec y_B$ and update the state of the remaining degrees of freedom according to the description offered in  Sec.~\ref{sec:continuousmonitor}, while 
using the augmented quantities $\vec {\tilde{a}}$, $\mat {\tilde{A}}$, and $\mat {\tilde{C}}$, in
$\vec {\tilde {m}} = \begin{psmallmatrix} \vec {\tilde{a}} \\ \vec{b} \end{psmallmatrix}$ and $\mat {\tilde\Gamma}= \begin{psmallmatrix} \mat{\tilde{A}} & \mat{\tilde{C}} \\ \mat{{\tilde{C}}}^T & \mat B \end{psmallmatrix}$, to simultaneously represent the variables $\vecop {\tilde y}_A = \begin{psmallmatrix} \vecop {{y}}_A \\ \vecop{f} \end{psmallmatrix}$ of the oscillators and the unknown perturbations $\vecop{f}$. The probe field variables $\vecop y_B$ retain their properties and dynamics and can be eliminated, which leads to effective equations of the form of \eqref{eqs:measurement} for the augmented mean vector $\vec{\tilde a}$ and covariance matrix  $\mat{ \tilde{A}}$. These objects explicitly provide the estimated value of $\vecop{f}$ and the variance of the estimate around the true value.


Our equations for $\vec{\tilde a}$ and  $\mat{ \tilde{A}}$ are equivalent to classical Kalman filter theory \cite{maybeck1979}, which was applied for magnetic field sensing by continuous QND probing of an atomic collective spin in Ref.~\cite{geremia2003quantum}, see also Ref.~\cite{molmer2004estimation,jimenez2018signal,amor2021noisy}. Our derivation of these equations from the ``quantized'' form of the unknown classical parameters provides a robust starting point to explore the sensitivity limits imposed by quantum mechanical uncertainty relations and measurement back action on the quantum meter system in more complex settings, see also Ref.~\cite{madsen2004spin}.    


Our analysis is developed for application with measurement data obtained in an experiment, but here we synthesize such data by the Algorithm~\ref{alg:synth} and subsequently we estimate the perturbation according to  Algorithm~\ref{alg}.  
\algrenewcommand{\alglinenumber}[1]{#1.}
\begin{algorithm}[H]
  \caption{\small Simulation of perturbations and measurement data }
  \label{alg:synth}
   \begin{algorithmic}[1]
   \State Numerically generate a particular classical realization of each perturbation $f_i(t)$ according to Eq.~\eqref{eq:OU}.
   \State Simulate the conditional dynamics for $\vec a(t)$ and $\mat A (t)$ of Sec.~\ref{sec:continuousmonitor} using $f_i(t)$ and store the detection record $\vec q\ut{meas}_B(t)$ from the numerical sampling procedure according to Eqs.~\eqref{eq:detectionrecord}.
    \end{algorithmic}
\end{algorithm}
\begin{algorithm}[H]
  \caption{\small Estimation of perturbations (filtering)}
  \label{alg}
   \begin{algorithmic}[1]
   \State Acquire a detection record, either from an  experiment or from a simulation by Algorithm~\ref{alg:synth}. 
   \State Use the acquired detection record to simulate the conditional dynamics for the augmented mean values $\vec {\tilde a}(t)$ and covariance matrix $\mat {\tilde{A}}(t)$.
   Definite elements of these  quantities provide the estimates of  $f_i(t)$ and their variances.
    \end{algorithmic}
\end{algorithm}

\subsection{Smoothing}
Algorithm~\ref{alg} gives an estimate for the perturbations at time $t$ conditioned upon all measurements until $t$.
Alternatively one might attempt to provide an estimate at time $t$ conditioned upon all measurements until and after $t$.
That is, given the time horizon $\mathcal H = [0,T]$ of the complete experiment, how can we benefit from \textit{all} measurements until time $T$ to retrodict the strength of the perturbations at any intermediate time $t \in \mathcal H$?


In classical estimation theory this question is answered by so-called smoothing filters, such as the  forward-backward or $\alpha/\beta$ filters for estimation of hidden Markov models \cite{rabiner1989a}, and the Mayne-Fraser-Potter two-filter smoother for Gaussian distributions \cite{mayne1966a,fraser1969the}. Quantum equivalents of these filters were first developed by Tsang in Refs.~\cite{tsang2009time,tsang2009optimal,tsang2010optimal}, and following the  previous sections, we shall present a self-contained derivation with reference to quantum measurement theory and the past quantum state (PQS) retrodiction formalism \cite{gammelmark2013past}. We shall apply smoothing to obtain the maximum achievable information from the ``negative mass" detection scheme in Fig.1. 

\subsubsection{The past quantum state}
The PQS theory is a generalization of the two-state formalism by Watanabe \cite{watanabe1955symmetry} and Aharonov et. al. \cite{aharonov1965time,reznik1995time} which was further developed in Ref.~\cite{gammelmark2013past} to retrodict the outcome probabilities for the unknown outcomes of past measurements on open and monitored quantum systems. Applying the PQS theory to our ``quantized" unknown classical variables $\vec{f}$, it becomes the Bayesian estimate of their values at time $t$, conditioned on all measurements, prior and posterior to $t$.
We shall briefly recall the derivation of the Past Quantum State at the level of Hilbert space operators before moving to the convenient Gaussian Wigner function representation of states and operators. 

The conditional dynamics due to the continuous monitoring of a quantum system up until time $t = n \dt$ is described by the unnormalized density matrix following application of a sequence of operators
\begin{align}\label{eq:povm1}
    \op \rho(t)
    =  \op \Omega_{\vec{q}_{n}} \dots \op \Omega_{\vec{q}_1} \op \rho(0) \op \Omega_{\vec{q}_1}^\dagger \dots \op \Omega_{\vec{q}_{n}}^\dagger.
\end{align}
Here, we assume for simplicity of notation that the system is subject to deterministic, unitary evolution and measurements with a sequence of outcome values $\vec{q}_i$ \eqref{eq:detectionrecord}, so that for each time interval the evolution is governed by a unitary operator $\op U$ and back action governed by POVM operators $\op M_{\vec{q}_i}$ \cite{nielsen2019quantum}, and $\op \Omega_{\vec{q}_i} = \op M_{\vec{q}_i} \op U$. In our setting, the projective measurements of a quadrature component of the probe field segments are dominated by their Gaussian fluctuations and cause  minute stochastic changes of the state of the observed oscillators, represented by the operators $\op M_{\vec{q}_i} = \op M_{({q_{n_B}})_i} \dots \op M_{({q_1})_i}$. 
The POVM operators are normalized as 
$\int_{} \op M_{\vec{q}}^\dagger \op M_{\vec{q}} \, d\vec q = \op\I $, and according to quantum measurement theory, the trace norm of \eqref{eq:povm1} yields the joint probability of all the specified outcome results.

Proceeding with a projective measurement $\op \Pi_{y_j}= \op\Pi_{y_j}^\dagger = \ket{y_j}\bra{y_j}$ at $t$ on the observed quantum system, followed by continued optical probing until $T=N \dt$ yields the conditioned state
\begin{align}
    \op \rho(T)
    = \op \Omega_{\vec{q}_{N}} \dots \op\Omega_{\vec{q}_{n+1}} \op\Pi_{y_j}\op\rho(t) \op\Pi_{y_j}^\dagger \op\Omega_{\vec{q}_{n+1}}^\dagger\dots \op \Omega_{\vec{q}_{N}}^\dagger,
\end{align}
and the joint probability density for the full optical detection record $\vec{q}_1, \dots,\vec{q}_N$
\textit{and} the projective outcome value $y_j$ at time $t$ is given by
\begin{align}
    P&(\vec{q}_1, \dots, y_j,\dots \vec{q}_N) = \tr[\op \rho(T)]  \notag \\
    & =  \tr[\op \Omega_{\vec{q}_{N}} \dots  \op \Omega_{\vec{q}_{n+1}} \op\Pi_{y_j}  \op\rho(t) \op\Pi_{y_j}^\dagger \op \Omega_{\vec{q}_{n+1}}^\dagger  \dots  \op \Omega_{\vec{q}_{N}}^\dagger] \notag \\
    &\equiv \tr[\op\Pi_{y_j} \op \rho(t)  \op\Pi_{y_j}^\dagger \op E(t) ], \label{eq:jointprobability}
\end{align}
where we used the cyclic property of the trace to define the so-called \textit{measurement effect operator} on the same Hilbert space as $\op \rho$.
\begin{align}
    \op E(t) = \op \Omega_{\vec{q}_{n+1}}^\dagger\dots \op \Omega_{\vec{q}_{N}}^\dagger \op \I \op \Omega_{\vec{q}_{N}} \dots \op \Omega_{\vec{q}_{n+1}}.
\end{align}
The expression for $\op E(t)$ is similar to the expression for the time evolved density matrix, except the operator $\op E(t)$ is conditioned on the detection record for times after $t$ and is found by a sequential backward evolution from the final value, $\op E(T)=\op \I$ by the adjoint of the POVM operators. For generalization to cases including damping and dissipation, see Ref.~\cite{gammelmark2013past}.

After all the field measurements have been done, the value of the joint probability distribution \eqref{eq:jointprobability} evaluated at the fixed entries of the detection record yields the (conditonal) probability for the still unknown outcome $y_j$. Conditioned on only measurements until time $t$, and on measurements before and after time $t$, this yields, 
\begin{subequations}
\label{eqs:probabilitites}
\begin{align}
    P(y_j | \vec{q}_{1} \dots \vec{q}_{n}) \hspace*{0.09cm}&\propto \tr[\op \Pi_{y_j} \op \rho(t)],  \\
    P\st{pqs}(y_j | \vec{q}_{1} \dots \vec{q}_{N}) &\propto \tr[\op \Pi_{y_j} \op \rho(t) \op \Pi_{y_j}^\dagger \op E(t)]  \label{eq:Ppqs}
\end{align}
\end{subequations}
respectively, where the probabilities are normalized by the sum (or integral) of the expressions over the argument $y_j$. 

The first expression is the usual Born rule, and it applies also when $\op \Pi_{y_j}$ represents the unknown classical perturbations, in which case the conditional quantum state $\op \rho(t)$ acts as a Bayesian \textit{filter}. The second expression retrodicts past measurement outcome probabilities \cite{gammelmark2013past}, and the pair of operators $\op \rho(t)$ and $\op E(t)$ corresponds to a Bayesian \textit{two-way smoothing filter} for the past value of the unknown classical perturbations \cite{tsang2009time, tsang2009optimal}. 

\subsubsection{Gaussian states}

To apply the Gaussian state formalism, we represent the augmented operator $\op {\tilde {E}}(t)$ 
in a manner similar to  Eq.~\eqref{eq:rhotilde} for $\op{\tilde{\rho}}$ \footnote{In the continuous basis, 
\begin{align*}
    \op {\tilde {E}} = \int d\vec x d\vec x' d\vec f \, \tilde E_{\vec f }^{\vec x \vec x'} \ket{\vec f }\bra{\vec f} \otimes \ket{\vec x}\bra{\vec x'}.
\end{align*}
}, 
\begin{align}
    \op{\tilde{E}} = \int d\vec f\, \ket{\vec f} \bra{\vec f} \otimes \op{E}_{\vec f}. 
\end{align}
Both $\op{\tilde{\rho}}$ and $\op{\tilde{E}}$ are operators on the reduced Hilbert space of the oscillators and perturbations, described by variables $\vecop {\tilde y}_A = \begin{psmallmatrix} \vecop {{y}}_A \\ \vecop{f} \end{psmallmatrix}$, and they   are evolved with Gaussian Wigner representations $\mathcal W_{\op {\tilde{\rho}}}^{\vec {\tilde{a}}_{\rho}, \mat {\tilde{A}}_{ \rho}}(\vec{\tilde{y}}_A)$
and $\mathcal W_{\op {\tilde{E}}}^{\vec {\tilde{a}}_{E}, \mat {\tilde{A}}_E}(\vec{\tilde{y}}_A)$ which are fully determined by the first and second moments $\mat{\tilde{A}}_\rho, \vec {\tilde{a}}_{\rho}$ and $\mat{\tilde{A}}_E, \vec {\tilde{a}}_{E}$.
The moments $\mat{\tilde{A}}_\rho, \vec {\tilde{a}}_{\rho}$ evolve according to the theory presented in the previous section and are now explicitly labelled with the index $\rho$ while first and second moments determined by the measurement effect operator follow a backward time evolution. For the evolution in each time step including the field segments, we have
\begin{subequations}
\begin{align}
    \vec {\tilde m}_E(t-\dt) &= \mat{\tilde{ D}}^-_{\dt}\mat {\tilde S}_{\dt}^- \vec {\tilde m}_E(t), \\
    \mat {\tilde\Gamma}_E(t-\dt) &= \mat{\tilde{ D}}_{\dt}^{-}\mat {\tilde S}_{\dt}^- \mat {\tilde \Gamma}_E(t) \mat {\tilde S}_{\dt}^{-T}\mat{\tilde{ D}}_{\dt}^{-T} + \mat {\tilde L}_{\dt}^{-}. 
\end{align}
\end{subequations}
The elements of $\mat{\tilde {S}}_{\dt}^-$ follow from the equation of motion $\op y_i(t-\dt) \approx \op y_i(t) - \dot{\op{y}}_i(t) \dt$.
When ``played in reverse", the OU process experiences negative damping but unchanged stochastic fluctuations---the matrix $\mat{\tilde{D}}_{\dt}^-$ is therefore obtained by negating $\gamma_i$ in $\mat{\tilde {D}}_{\dt}$ while the stochastic diffusion rates $\sigma_i$ are unchanged and $\mat{\tilde {L}}_{\dt}^- = \mat{\tilde {L}}_{\dt}$ (see details in Sec.~\ref{sec:physical} for the setup in Fig.~\ref{fig:setup}). 

The monitoring of the field components leads to a stochastic sequence of POVM operators, which shall be applied in reverse to yield $\op {\tilde{E}}$. The effective measurement updates of $\vec{\tilde{a}}_{E}(t)$ and $\mat{\tilde{A}}_{E}^{}(T)$ governed by the detection record $\vec q_B\ut{meas}(t)$ are thus on the same form as Eq.~\eqref{eqs:measurement}.   
The Wigner function for $\op E(T) = \op \I$ is a uniform Gaussian distribution (the constant function with infinite variance) on the oscillators and classically unknown parameters. In practice we assume vanishing mean values and 
$\mat{\tilde{A}}_{E}^{}(T) = \text{diag}(\dots, v, \dots)$ 
with a suitably large dimensionless $v = 10^{4}$ at $t=T$, and we have verified that the backward evolved first and second moments quickly  become independent of these values.

Rather than explicitly reconstructing the operators $\op {\tilde{\rho}}$ and $\op {\tilde{E}}$, it is much more convenient for our purpose to refer to their first and second moments and their corresponding Gaussian Wigner functions.
This is because the trace of a product of two operators equals the integral of the product of their Wigner functions \eqref{eq:wigner_expectation}, and the PQS expression, $P\st{pqs}(f_j')  \propto \tr\left[ \Big( \ket{f_j'} \bra{f_j'} \op{\tilde{ \rho}} \Big) \Big(\ket{ f_j'} \bra{f_j'} \op{\tilde{E}}\Big) \right]$ for the variable $y_j \rightarrow f_j'$ in Eq.~\eqref{eq:Ppqs} involves such a product of operators $\ket{f_j'}\bra{f_j'}\op{\tilde{X}}$, with $\, X = \rho, E$. 

Multiplication of $\op{\tilde{\rho}}$ and $\op{\tilde{E}}$ by the projection operator yields the transformed Wigner function \cite{supplementary}
\begin{align}
\mathcal W_{\ket{f_j'}\bra{f_j'}\op{\tilde{X}}} (\vec{\tilde y}_A)= \delta(f_j - f_j')\mathcal W_{\op{\tilde{X}}} (\vec{\tilde y}_A), 
\end{align}
and the PQS probability density for $f_j'$ in Eq.~\eqref{eq:Ppqs} can be directly written in terms of the Gaussian Wigner functions, 
\begin{align}
    P\st{pqs}(f_j') & \propto \tr\left[ \Big(\ket{f_j'} \bra{f_j'} \op{\tilde{ \rho}}\Big)\Big( \ket{ f_j'} \bra{f_j'} \op{\tilde{E}} \Big)\right]  \notag \\
    &\propto \int  d\vec x_A d\vec p_A d\vec f\; \mathcal W_{\ket{f_j'}\bra{f_j'}\op{\tilde{\rho}}} \times \mathcal W_{\ket{f_j'}\bra{f_j'}\op{\tilde{E}}}  \notag \\
    &= \int  d\vec x_A d\vec p_A d\vec f_{i\neq j} \; \big( \mathcal  W_{\op{\tilde{\rho}}}^{\vec {\tilde{a}}_{\rho}, \mat {\tilde{A}}_{ \rho}}\!\! \times \mathcal W_{\op{\tilde{E}}}^{\vec {\tilde{a}}_{E}, \mat {\tilde{A}}_{ E}}\big)|_{f_j=f_j'} \notag \\
    &= \int  d\vec x_A d\vec p_A d\vec f_{i\neq j} \; \mathcal{W}_{\op {\tilde{\rho}}, \op {\tilde{E}}}^{\vec {\tilde{a}}_{\rho,E}, \mat {\tilde{A}}_{ \rho,E}}(\vec y_A, \vec f_{i\neq j}, f_j' ). 
    \label{eq:pqs_gaussian}
\end{align}
The integrals in the last two lines are over all variables $\vec{\tilde{y}}_A$ except the desired value of the argument $f_j'$.
In the last expression we have used that the product of two multivariate Gaussian functions is also a multivariate Gaussian function, 
with the covariance matrix and mean values given by 
\begin{subequations}
\label{eqs:PQS_moments}
\begin{align}
    \mat{\tilde{A}}_{\rho, E}^{-1} &= \mat{\tilde{A}}_{\rho}^{-1} + \mat{\tilde{A}}_{E}^{-1}, \\
    \vec{\tilde{a}}_{\rho, E} &= \mat{\tilde{A}}_{\rho, E}\left( \mat{\tilde{A}}_{\rho}^{-1} \vec{\tilde{a}}_{\rho} + \mat{\tilde{A}}_{E}^{-1} \vec{\tilde{a}}_{E}\right).
\end{align}
\end{subequations}
These expressions directly yield the retrodicted (smoothed) estimate of the time dependent perturbations and their accompanying variances based on the entire detection record. 
The implementation of the formalism is summarized in Algorithm~\ref{alg2}.

\algrenewcommand{\alglinenumber}[1]{#1.}
\begin{algorithm}[H]
  \caption{\small Estimation of the perturbations by smoothing}
  \label{alg2}
   \begin{algorithmic}[1]
   \State Obtain data record from experiments or by simulation with Algorithm \ref{alg:synth}.
   \State Apply Algorithm~\ref{alg} to obtain the conditional dynamics for  $\vec{\tilde{a}}_{\rho}(t)$ and $\mat{\tilde{A}}_{\rho}^{}(t)$. 
   \State Use the same measured or simulated detection records to obtain the backward conditional dynamics for $\vec {\tilde a}_{E}(t)$ and $\mat{\tilde{A}}_E(t)$. 
   \State Calculate the retrodicted PQS moments $\vec{\tilde{a}}_{\rho, E}(t)$ and $\mat{\tilde{A}}_{\rho, E}^{}(t)$ in Eq.~\eqref{eqs:PQS_moments}.
   Definite elements of these  quantities provide the smoothed, PQS estimates of  $f_i(t)$ and their variances.
    \end{algorithmic}
\end{algorithm}
%


\section{Numerical examples}
\label{sec:numerical}
We have presented a general, compact mathematical formalism for the Gaussian quantum states and estimates of classical parameters conditioned on continuous quadrature (homodyne) measurements on  probe fields. We shall now demonstrate numerical application of the formalism and evaluate the assessment of the probing scenario  illustrated in Fig.~\ref{fig:setup} and described by Eq.~\eqref{eq:hamiltonian}. The entanglement and noise cancellation properties of the scheme have been analyzed and experimentally demonstrated in the frequency domain \cite{thomas2021entanglement}, i.e., noise power spectra of the probe signal measurements have been compared and agree with theory. Here, we shall address the explicit time domain analysis, and provide the time dependent estimator and compare it with its true value in simulated experiments.   

\subsection{Physical parameters}
\label{sec:physical}

We assume the Hamiltonian in Eq.~\eqref{eq:hamiltonian}, describing how the oscillator variables $\vecop y_A=(\x[S_1], \p[S_1], \x[S_2], \p[S_2])$ interact sequentially with the field variables of the consecutive light segments $\vecop y_B=(\x[L_1], \p[L_1], \x[L_2], \p[L_2])$. 

By assuming fewer degrees of freedom and vanishing values of some of the interaction parameters in Eq.~\eqref{eq:hamiltonian} we can use the same theoretical model to study the estimation of single or multiple perturbations, by a single or two probe beams and a single or two oscillator modes. We can thus directly observe how the back action evading mechanism  affects the dynamics of the physical systems and the  estimation of the perturbations. 

We henceforth assume for convenience that all physical parameters are given in units of the values listed in Table~\ref{tab:parametervalues} applicable, e.g., to the probing of ensemble atomic spins, subject to magnetic field fluctuations \cite{molmer2004estimation,cheng2020estimating}. The matrix and vector quantities applied in our  formalism are listed in the following equations,

\onecolumngrid

\begin{widetext}
    \begin{align}
        \setcounter{MaxMatrixCols}{11}
        \mat {\tilde{S}}_{\dt} &= 
        \begin{pNiceMatrix}[first-row,last-col]
                     {x_{A_1}}   & {p_{A_1}}   & {x_{A_2}}   & {p_{A_2}}   & {f_x}   & {f_p}   & {x_{L_1}}   & {p_{L_1}}   & {x_{L_2}}   & {p_{L_2}} \\
         \cos\omega_{1}\dt      & \sin \omega_{1}\dt     & 0                     & 0                     & c_{x}\dt           & 0                 & 0                     & \kappa_{1,1}\sqrt{\dt}        & 0                     & 0  & x_{A_1}    \\
         -\sin\omega_{1}\dt     & \cos \omega_{1}\dt     & 0                     & 0                     & 0                 & c_{p}\dt           & 0                     & 0                     & -\kappa_{1,2}\sqrt{\dt}       & 0  & p_{A_1}    \\ 
         0                     & 0                     & \cos\omega_{2}\dt      & \sin\omega_{2}\dt      & 0                 & 0                 & 0                     & \kappa_{2,1}\sqrt{\dt}        & 0                     & 0  & x_{A_2}    \\
         0                     & 0                     & -\sin\omega_{2}\dt     & \cos\omega_{2}\dt      & 0                 & 0                 & 0                     & 0                     & -\kappa_{2,2}\sqrt{\dt}       & 0  & p_{A_2}    \\
         0                     & 0                     & 0                     & 0                     & 1    & 0                 & 0                     & 0                     & 0                     & 0  & f_x        \\
         0                     & 0                     & 0                     & 0                     & 0                 & 1    & 0                     & 0                     & 0                     & 0  & f_p        \\
         0                     & \kappa_{1,1}\sqrt{\dt}        & 0                     & \kappa_{2,1}\sqrt{\dt}        & 0                 & 0                 & 1                     & 0                     & 0                     & 0  & x_{L_1}    \\
         0                     & 0                     & 0                     & 0                     & 0                 & 0                 & 0                     & 1                     & 0                     & 0  & p_{L_1}    \\
         0                     & 0                     & 0                     & 0                     & 0                 & 0                 & 0                     & 0                     & 1                     & 0  & x_{L_2}    \\
         -\kappa_{1,2}\sqrt{\dt}       & 0                     & -\kappa_{2,2}\sqrt{\dt}       & 0                     & 0                 & 0                 & 0                     & 0                     & 0                     & 1  & p_{L_2}    
    \end{pNiceMatrix}, \label{eq:Stilde}\\
    \mat {\tilde{D}}_{\dt}  &= \text{diag}(1,1,1,1,(1-\gamma_x\dt), (1-\gamma_p\dt),1,1,1,1),\label{eq:Dtilde}\\
    \mat {\tilde{L}}_{\dt} &= \text{diag}(0,0,0,0, 2\sigma_x \dt, 2\sigma_p\dt, 0,0,0,0). \label{eq:Ltilde} \\
        \vec F_{\dt} &= (c_x f_x \dt, c_p f_p \dt, 0, 0, 0,0,0,0)^T, \label{eq:F}\\
        \mat \Pi &= \mat \pi_{L_1} \oplus \mat \pi_{L_2} = \mat \pi_x \oplus \mat \pi_p = 
    \text{diag}(1, 0, 0, 1), \label{eq:Pi} \\
    \mat \Delta_B\ut{meas}  &= \text{diag}\Big( \Delta_{L_1}, 0,0, \Delta_{L_2}\Big) \equiv \text{diag}\Big(x_{L_1}\ut{meas} - \braket{\x_{L_1}}, 0,0, p_{L_2}\ut{meas} - \braket{\p_{L_2}}\Big), \label{eq:DeltaBmeas}
    \end{align}
\end{widetext}

\twocolumngrid

When the classical perturbations have known values, they displace the quadrature observables $\vecop y = \begin{psmallmatrix} \vecop y_A \\ \vecop y_B \end{psmallmatrix}$ by the argument $\vec F_{\dt}$ in Eq.~\eqref{eq:F}, and the interaction matrix $\mat S_{\dt}$ for the quantum variables is given for small values of  $\dt$ by omitting the $f_x$ and $f_p$ rows and columns in $\mat {\tilde {S}}_{\dt}$ in Eq.~\eqref{eq:Stilde}. 
When the perturbing  fields are instead represented as ancillary quantum variables, i.e. $f_i \rightarrow \op f_i$ in $\H$, the augmented system variables are
\begin{align*}
\vecop {\tilde {y}} = \begin{pmatrix} \vecop {\tilde{y}}_A \\ \vecop y_B \end{pmatrix}
\quad \text{where}\quad 
\vecop {\tilde {y}}_A = \begin{pmatrix}\vecop y_A \\ \op f_x \\ \op f_p\end{pmatrix},
\end{align*}
and the full $\mat {\tilde{S}}_{\dt}$ matrix and OU dissipation $\mat{\tilde{D}}_{\dt}$ in Eq.~\eqref{eq:Dtilde}, and OU diffusion $\mat {\tilde {L}}_{\dt}$ in Eq.~\eqref{eq:Ltilde} are used.


Rather than extensively scoping out the dependence of the results on different combinations of physical parameters, we focus our numerical efforts on highlighting the behavior with few fixed parameter settings, specified by the values in Table~\ref{tab:parametervalues}, and observing how the switching on and off of different terms change the sensing mechanisms and hence performance.



\clearpage

\onecolumngrid

\newcommand{\thetoptrimVar}{+0.3}
\newcommand{\thebottrimWigner}{-0.17}
\newcommand{\thetoptrimWigner}{+0.20}
\newcommand{\thelefttrimWigner}{-0.0}
\newcommand{\theleftspaceWigner}{0}

\begin{figure}[t]
    %
    \includegraphics[]
    {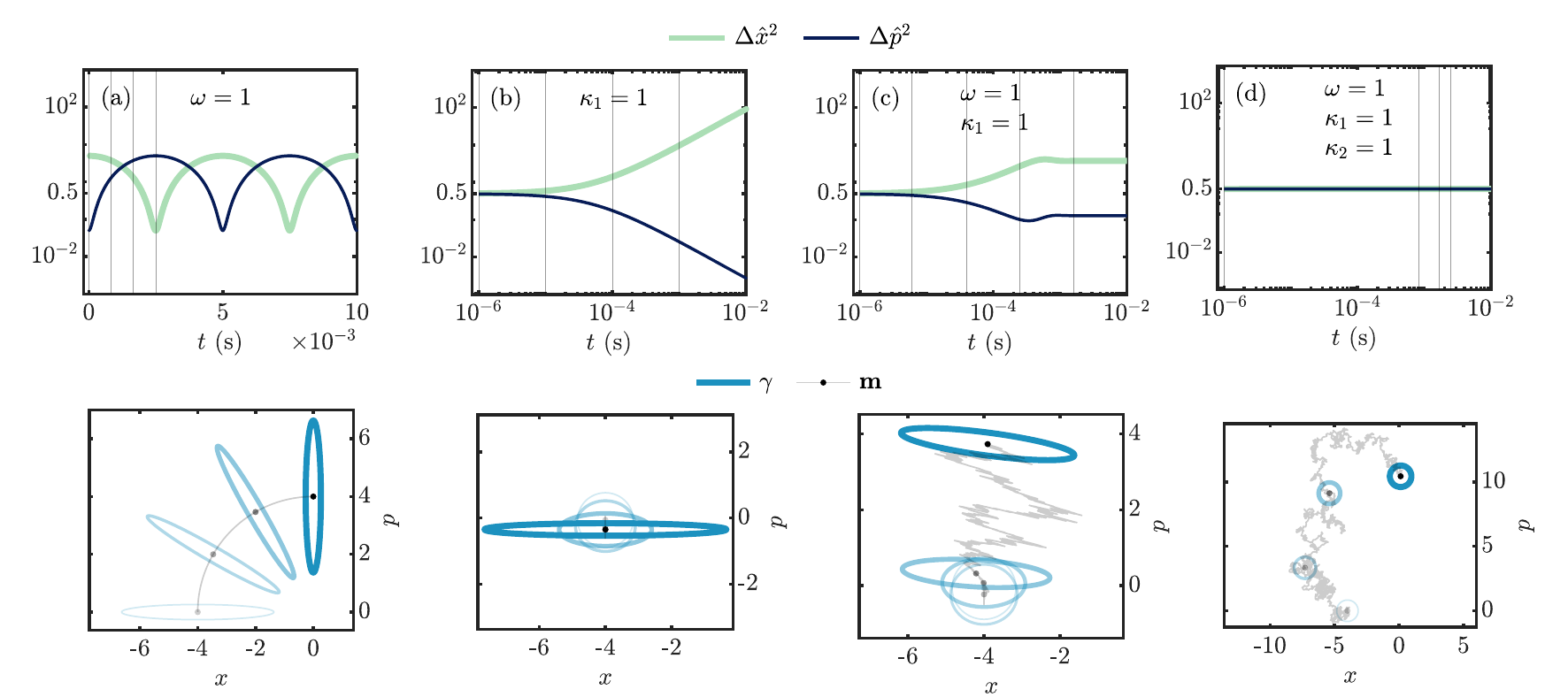}
    \caption{ 
    Evolution of a single oscillator. Numerical examples for different scenarios (columns) with nonzero parameter values from Table~\ref{tab:parametervalues}. 
    (a): Squeezed state oscillating without measurements. 
    (b): Coherent state squeezed by continuous optical measurements sensitive to the oscillator quadrature $\p$ . 
    (c): As (b) but including oscillations. 
    (d): As (c) but including continuous optical measurements sensitive to the oscillator quadrature $\x$. 
    Variances are shown as function of time in the upper panels. Snapshots of  associated Wigner function contour lines are shown in the lower panels around the time dependent mean values (translucent grey solid line), as in Fig.~\ref{fig:WignerExample}. 
    The snapshot times are indicated by vertical lines in the upper panels and their sequence is indicated by increasing line width and opacity. 
    }
    \label{fig:fieldfree_1osc}
\end{figure}

\twocolumngrid

\begin{table}[]
    \centering
    \begin{tabular}{l l }
        \textbf{Parameter} & \textbf{Value} \\
        \hline
         Time step & $\dt = 10^{-5}-10^{-7} \,\si{s}$ \\
         Oscillation frequencies & $|\omega_i| = 0.1 \times 2\pi/\si{ms} \approx 0.63 \, \si{kHz}$ \\
         Light couplings & $|\kappa_{i,j}| =  135 \,\si{Hz}$ \\         
         Classical field couplings & $c_x = c_p =  1.5 \times  10^4 \,\si{Hz}^{}$ \\
         OU damping rate of $f_x$ & $\gamma_x =  100 \,\si{Hz}$ \\
         OU diffusion rate of $f_x$  & $\sigma_x = 10  \,\si{Hz}$ \\
         OU damping rate of $f_p$ & $\gamma_p =  10 \,\si{Hz}$ \\
         OU diffusion rate of $f_p$  & $\sigma_p = 1  \,\si{Hz}$ 
    \end{tabular}
    \caption{Parameter values used, if not set to zero, in the numerical simulations ($i,j=1,2$). These values may for example represent vector magnetometry with atomic spin ensembles.}
    \label{tab:parametervalues}
\end{table}

\subsection{Monitoring of unperturbed oscillators}
In this section we consider scenarios where the classical perturbations are absent $f_x(t) = f_p(t) = 0$ or, equivalently, $c_x = c_p = 0$.

\subsubsection{Probing of a single oscillator}
Consider a single mode $\vecop y=\begin{psmallmatrix} \x \\ \p \end{psmallmatrix}$
with dimensionless  
canonical variables $[\x[], \p] = i$ with $\covar{}{\x} \covar{}{\p} \geq 1/4$ \footnote{For a mechanical oscillator $\H_\omega = \frac{p^2\st{mech}}{2m} + \frac{1}{2} m \omega^2 x^2\st{mech}$ with mass $m$ the dimensionless variables are given by $\x = \x\st{mech}/x_0$ and $\p = \p\st{mech} x_0/\hbar$ where $x_0^2 = \hbar/(m\omega)$.}, subject to the harmonic oscillator Hamiltonian with corresponding Heisenberg time evolution
\begin{subequations}
\label{eqs:harmonicosc}
\begin{align}
    \H_\omega &= \frac{\hbar\omega}{2}(\x[]^2 + \p^2), \\
    \x[](t) &= \x[](0) \cos\omega t + \p[](0) \sin\omega t ,  \\
    \p[](t) &= \p[](0) \cos\omega t - \x[](0) \sin\omega t. 
\end{align}
\end{subequations}
In suitable interaction pictures (rotating frames), systems may be described by positive, negative and vanishing values of $\omega=0$ in (\ref{eqs:harmonicosc}). 
The back action from a precise  measurement of $\x$ at $t=0$ squeezes the variance $\covar{}{\x(0)} \rightarrow 0$ and induces a corresponding anti-squeezing $\covar{}{\p(0)} \rightarrow \infty$.
For $\omega\neq 0$, the reduced and increased variances will then oscillate back and forth between the two operators as their time evolution amounts to rotation about the origin in phase space with angular frequency $\omega$ as shown in Fig.~\ref{fig:fieldfree_1osc}(a). 

Rather than such an abrupt  measurement and preparation of a squeezed state, we consider continuous probing by the sequential interaction with infinitesimal segments of a laser probe beam with quantum degrees of freedom $\vecop y_B = (\x[L], \p[L])^T$ subject to a subsequent measurement. 
Specifically, an interaction with $\p[L]$ and measurement of $\x[L]$
\begin{align}
    \H_\kappa^p = \hbar \kappa \p[S]\p[L], \quad \mat \pi_1 = \mat \pi_x,
\end{align}
results in a small incremental squeezing of $\p[S]$ and anti-squeezing of $\x[S]$ while 
\begin{subequations}
\begin{align}
    \H_\kappa^x = \hbar \kappa \x[S]\x[L], \quad \mat \pi_1 = \mat \pi_p,
\end{align}
results in a small incremental squeezing of $\x[S]$ and anti-squeezing of $\p[S]$.
\end{subequations}
This leads to the following distinct scenarios and accumulated effects on the oscillator. 

\begin{itemize}[itemindent=-0.25 em,align=left,  leftmargin=*]
    \item[\textit{Single probe}, {$\omega = 0$}] --- squeezing accumulates with a variance scaling asymptotically as $1/t$ [Fig.~\ref{fig:fieldfree_1osc}(b)]. 
    \item[\textit{Single probe}, {$\omega \neq 0$}] --- squeezing is partially counteracted by the rotation and the variances become asymptotically constant [Fig.~\ref{fig:fieldfree_1osc}(c)].
    \item[\textit{Two probes}, $\omega=0$ \textit{or} {$\omega \neq 0$}] --- two light modes $\vecop y_B = (\x[L_1], \p[L_1], \x[L_2], \p[L_2])^T$ interacting with different quadratures 
        \begin{align}
            \H^{}_{\kappa_1,\kappa_2} = \H^{p}_{\kappa_1} + \H^{x}_{\kappa_2}, \quad \begin{matrix} \mat \pi_1 = \mat \pi_x \\ \mat \pi_2 = \mat \pi_p \end{matrix}, \label{eq:1osc2lightH}
        \end{align}
        the squeezing effects compete, and for $\kappa_1 = \kappa_2$ no net squeezing occurs [Fig.~\ref{fig:fieldfree_1osc}(d)].
\end{itemize}

\subsubsection{Probing of two oscillators --- EPR variables}
Consider now a two-mode system $\vecop y_A = (\x[S_1], \p[S_1], \x[S_2], \p[S_2])^T$ with $[\x[S_i], \p[S_j]] = i \delta_{i,j}$ for $i,j=1,2$ 
and a separable oscillator Hamiltonian 
\begin{align}
    \H_{\omega_1, \omega_2} = \H_{\omega_1} + \H_{\omega_2}.
\end{align}
Precise knowledge of both quadratures for the individual oscillators is prohibited by the Heisenberg uncertainty relation, 
yet we can define four new EPR-type position and momentum variables \cite{einstein1935can},
\begin{align}
\vecop y^\pm_A \equiv \begin{pmatrix} \x[-] \\ \p[+] \\ \x[+] \\ \p[-]\end{pmatrix} & \equiv  
\frac{1}{\sqrt{2}} \begin{pmatrix} \x[S_1] - \x[S_2] \\ \p[S_1] + \p[S_2] \\ \x[S_1] + \x[S_2] \\ \p[S_1] - \p[S_2]\end{pmatrix}, 
\end{align}
satisfying the commutation relations
\begin{align}
    [\x[-], \p[+]] &= [\x[+], \p[-]] = 0, \\
    [\x[-], \p[-]] &= [\x[+], \p[+]] = i.
\end{align}
Hence $\x[-]$ and $\p[+]$ can both be principally known to arbitrary precision, $\covar{}{\x[-]}\covar{}{\p[+]} \geq 0$, while the values of the adjoint variables $\x[+]$ and $\p[-]$ become correspondingly uncertain.
The EPR variables can be written $\vecop y^\pm_A= \mat\R \vecop y_A$ 
with the matrix
\begin{align}
\mat \R &= \frac{1}{\sqrt{2}}
\begin{pmatrix*}[r]
1 & 0 & -1 & 0 \\
0 & 1 & 0 & 1   \\
1 & 0 & 1 & 0 \\
0 & 1 & 0 & -1
\end{pmatrix*},
\end{align}
and their first and second moments are given by
\begin{subequations}
\begin{align}
    \vec a^\pm  = \mat \R\vec a, \qquad 
    \mat A^\pm = \mat \R \mat A \mat \R^T. 
\end{align}
\end{subequations}
Two laser probes $\vecop y_B = (\x[L_1], \p[L_1], \x[L_2], \p[L_2])^T$ permit simultaneous measurements of $\x[-]$ and $\p[+]$ if we assume the interactions 
\begin{align}
    \H^{p}_{\kappa_{1,1}, \kappa_{2,1}} 
    &= \hbar \left( \kappa_{1,1} \p[S_1]  + \kappa_{2,1} \p[S_2] \right) \p[L_1], \notag \\
    &(\text{assume }   \kappa_{1,1} = \kappa_{2,1} \equiv \kappa_p / \sqrt{2} ) \notag \\
    &= \hbar \kappa_p \p[+] \p[L_1] \equiv \H^{p_+}_{\kappa_{p_+}}, \quad \mat \pi_1 = \mat \pi_x, 
\end{align}
and   
\begin{align}
    \H^{x}_{\kappa_{1,2}, \kappa_{2,2}} 
    &= \hbar \left( \kappa_{1,2} \x[S_1]  +  \kappa_{2,2} \x[S_2] \right) \x[L_2], \notag \\
    &(\text{assume } \kappa_{1,2} = -\kappa_{2,2} \equiv \kappa_x/\sqrt{2} )\notag \\
    &= \hbar \kappa_x \x[-] \x[L_2] \equiv \H^{x_-}_{\kappa_{x_-}}, \quad \mat \pi_2 = \mat \pi_p.
\end{align}
In summary, we have
\begin{align}
    \H^{p_{+}}_{\kappa_{p}} + \H^{x_{-}}_{\kappa_{x}}  \equiv  \hbar \kappa_{p} \p[+] \p[L_1] + \hbar \kappa_{x} \x[-] \x[L_2] , \quad \begin{matrix} \mat \pi_1 = \mat \pi_x \\ \mat \pi_2 = \mat \pi_p\end{matrix},
\end{align}
which  is similar to Eq.~\eqref{eq:1osc2lightH}, but here the operators $\x[-]$ and $\p[+]$ commute and can be simultaneously measured and squeezed. 
This leads to the following distinct accomplishments. 


\begin{figure}[t]
    \centering
    \includegraphics[]                                          {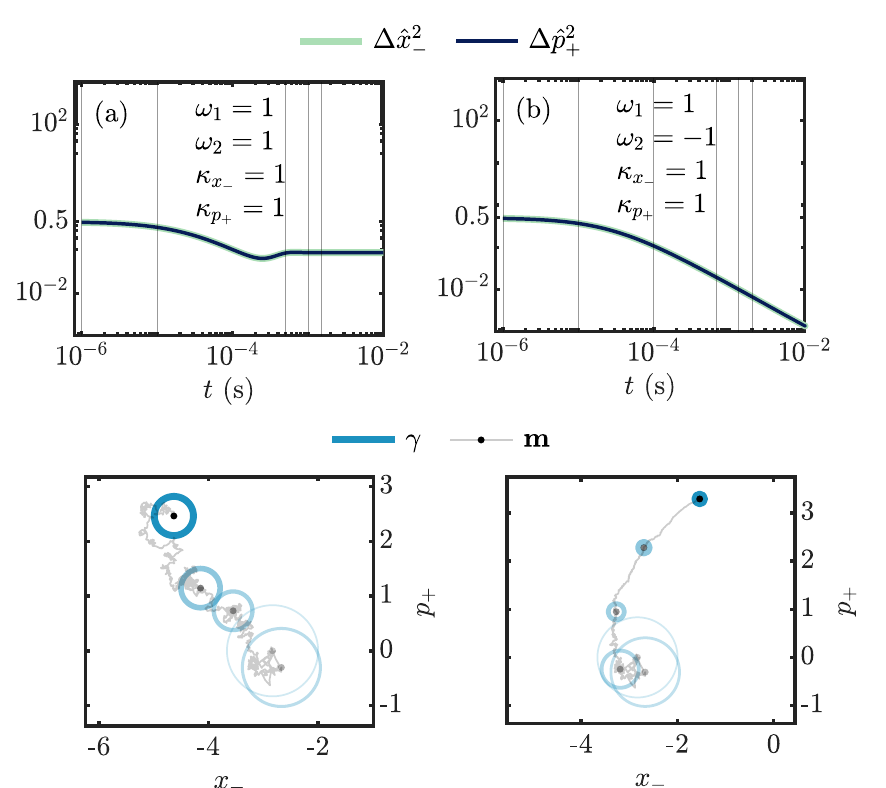} 
    \caption{Evolution of two oscillators. As Fig.~\ref{fig:fieldfree_1osc} but for commuting EPR variables $\x[-]$ and $\p[+]$ continuously measured indirectly starting from an initial coherent state. The oscillator frequency $\omega_1$ is positive and (a): $\omega_2$ is positive 
    or (b): $\omega_2$ is negative. The variance reduction in (b) is also attained for $\omega_1=\omega_2=0$.}
    \label{fig:fieldfree_2osc}
\end{figure}

\begin{itemize}[itemindent=-0.25 em,align=left,  leftmargin=*]
\item[\textit{Same-sign}, $\omega = \omega_1=\omega_2$] 
--- the free evolution of the oscillators yields
\begin{align}
    \begin{pmatrix}
    \x[-](t) \\
    \p[+](t)
    \end{pmatrix}
     &=
     \begin{pmatrix}
     \x[-](0) \cos\omega t + \p[-](0) \sin\omega t  \\
     \p[+](0) \cos\omega t - \x[+](0) \sin\omega t
     \end{pmatrix}, 
\end{align}
and the squeezed and anti-squeezed uncertainties oscillate periodically between the $\x[-]$ and $\p[-]$ variables (resp. $\p[+]$ and $\x[+]$), while the continuously probed system reaches a finite steady  squeezing of the collective observables, see. Fig.~\ref{fig:fieldfree_2osc}(a)
\item[\textit{Opposite-sign}, $\omega = \omega_1=-\omega_2$] --- the free evolution of the oscillators yields 
\begin{align}
    \begin{pmatrix}
    \x[-](t) \\
    \p[+](t)
    \end{pmatrix}
     &=
    \begin{pmatrix}
         \x[-](0) \cos\omega t + \p[+](0) \sin\omega t  \\
         \p[+](0) \cos\omega t - \x[-](0) \sin\omega t
    \end{pmatrix},
\end{align}
which show that the variables $\x[-]$ and $\p[+]$ do not couple to the antisqueezed observables and are both squeezed by the optical probing, see Fig.~\ref{fig:fieldfree_2osc}(b).
\end{itemize}

In conclusion, by choosing appropriate probe strengths and oscillator frequencies, $\kappa_{1,1} = \kappa_{1,2} = \kappa_{2,1} = -\kappa_{2,2}$, $ \omega_1 = -\omega_2$, the EPR variables $\x[-]$ and $\p[+]$ in Fig.~\ref{fig:setup} can both be determined to arbitrary precision with no adverse measurement back action or coupling to their conjugate, anti-squeezed observables.
As proposed by Tsang and Caves \cite{tsang2010coherent,tsang2012evading} this property can be leveraged for the sensing of perturbations acting on one of the oscillators. 


\subsection{Estimation of constant perturbations}
We now consider the estimation of the unknown value of constant perturbations ($\gamma_i=\sigma_i=0$) on the oscillator system. 
We assume initial oscillator ground states, ${a}_i = 0$ and ${A}_{ij} = \delta_{ij}$, and perturbations governed by an initial Gaussian distribution with vanishing mean and variance of 0.05 in all our simulations. 

It is possible to asymptotically approach the true values $\Braket{f_i} \rightarrow f_i$ to an arbitrary precision by measuring for longer and longer times, but the rate at which the variance of our estimate decreases depends on the measurement scenario
as shown in Fig.~\ref{fig:cstfield}.
All parameters are given in units of the values in Table~\ref{tab:parametervalues}.

For a single oscillator with $\omega_1=0$, subject to a single probe beam, it is possible to asymptotically achieve $\covar{}{f_p} \propto 1/t^3$ for the sensing of a single unknown  perturbation $f_p$, while we have no sensitivity to the value of $f_x$, see Fig.~\ref{fig:cstfield}(a). This reflects that the number of measurements (``samples") increases linearly with time, while the measurements are performed with a progressively better (more squeezed) sensor c.f. Fig.~\ref{fig:fieldfree_1osc}(b). 
The right panel shows the estimated and true values of $f_x$ and $f_p$ (assuming very distinct true values for ease of identification).

By coupling the oscillator quadratures to  two different perturbations and probe beams in Fig.~\ref{fig:cstfield}(b), the variance on the estimator of both perturbations approach zero but at slower asymptotic rates $\covar{}{f_x}, \covar{}{f_p} \propto 1/t$. This is because squeezing of both oscillator quadratures is prohibited by the uncertainty relation, and hence the sensor does not improve with time, c.f. Fig.~\ref{fig:fieldfree_1osc}(d).

We now introduce the second oscillator. 
Initially assuming $\omega_1 = \omega_2 = 0$ and probing the EPR variables $\x[-]$ and $\p[+]$ restores the benefits of squeezing c.f. Fig.~\ref{fig:fieldfree_2osc}(b) and results in an asymptotic $\covar{}{f_x},  \covar{}{f_x} \propto 1/t^3$ scaling for our estimate of both perturbations as shown in Fig.~\ref{fig:cstfield}(c). The right panel shows that the correct values of the perturbations are rapidly identified with high precision.

The lower panels in Fig.~\ref{fig:cstfield}(d) show the results for probing  EPR variables $\x[-]$ and $\p[+]$ of two oscillators with finite, opposite oscillator frequencies, $\omega_2 = -\omega_1$. 
These are variables are both squeezed, c.f., Fig.~\ref{fig:fieldfree_2osc}(b), and hence we observe the accelerated estimation of both $f_x$ and $f_p$, by $\covar{}{f_i} \propto 1/t^3$.  For longer times, however, we observe a cross over to a  $\covar{}{f_i} \approx \omega^2 / (2 c_i^2 \kappa^2 t)$ asymptotic dependence, which is analyzed further in \cite{supplementary}.

In practice, the asymptotic behaviors shown in Fig.~\ref{fig:cstfield} are subject to further change when damping and decoherence are taken into account \cite{madsen2004spin}. In this article we are striving to estimate temporally varying fluctuations, and  they will anyway reach constant asymptotic variances.   


\newcommand{\thetoptrim}{+0.98}
\newcommand{\thebottrim}{+0.7}
\begin{figure}
    \centering
    \includegraphics[trim={0cm \thebottrim cm 0cm 0cm}, clip]               {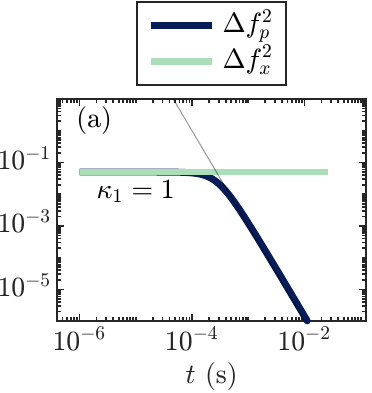}
    \includegraphics[trim={0cm \thebottrim cm 0cm 0cm}, clip]               {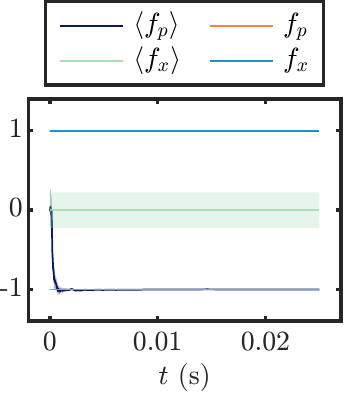} \\
    \includegraphics[trim={0cm \thebottrim cm 0cm \thetoptrim cm}, clip]    {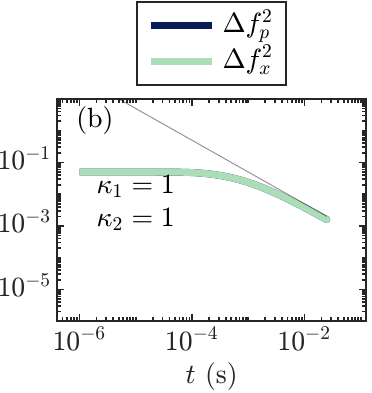}
    \includegraphics[trim={0cm \thebottrim cm 0cm \thetoptrim cm}, clip]    {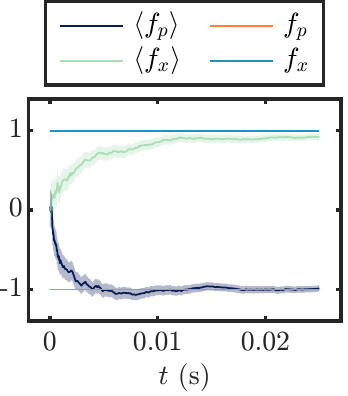} \\
    \includegraphics[trim={0cm \thebottrim cm 0cm \thetoptrim cm}, clip]    {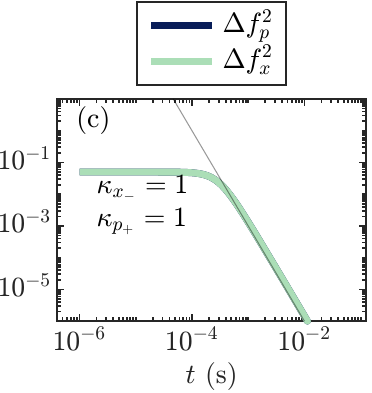}
    \includegraphics[trim={0cm \thebottrim cm 0cm \thetoptrim cm}, clip]    {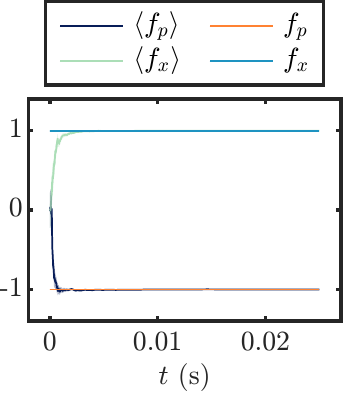} \\
    \includegraphics[trim={0cm  0cm 0cm \thetoptrim cm}, clip]              {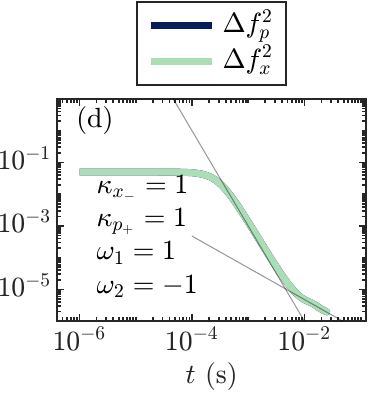} 
    \includegraphics[trim={0cm  0cm 0cm \thetoptrim cm}, clip]              {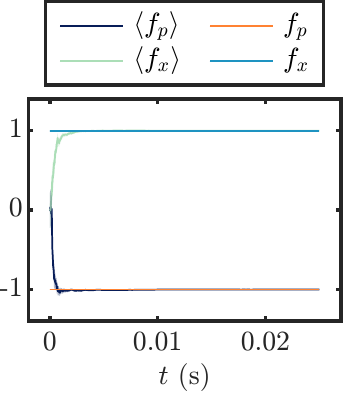} 
    \caption{
    Estimation of constant perturbations ($\gamma_i=\sigma_i=0$). 
    The left panels show the variance $\covar{}{f_i}$ with $1/t$ and  $1/t^3$ trend lines. The right panels show the true and estimated perturbation $\Braket{f_i} \pm  \std{}{f_i}$ (the true values are taken to be large positive and negative to distinguish their estimates clearly in the panels).  
    The upper panels (a) show the results for probing of $\p[S_1]$ for a single oscillator with $\omega_1=0$ being sensitive to only the value of $f_p$. 
    Panels (b) show the results for probing of $\x[S_1]$ and $\p[S_1]$ of  a single oscillator with $\omega_1=0$, being sensitive to both $f_x$ and $f_p$, but without the benefit of squeezing of the oscillator. 
    Panels (c) are for the probing and squeezing of EPR variables $\x[-]$ and $\p[+]$ of two oscillators with $\omega_1=\omega_2=0$, and hence the accelerated estimation of both $f_x$ and $f_p$. 
    The lower panels (d) are for the probing  of EPR variables $\x[-]$ and $\p[+]$ of two oscillators with opposite, non-vanishing frequencies. All parameters are given in units of the values in Table~\ref{tab:parametervalues}.
    }
    \label{fig:cstfield}
    \end{figure}
    

\subsection{Estimation of fluctuating perturbations}
\begin{figure}[h]
    \centering
    \includegraphics[trim={0cm \thebottrim cm 0cm 0cm}, clip]               {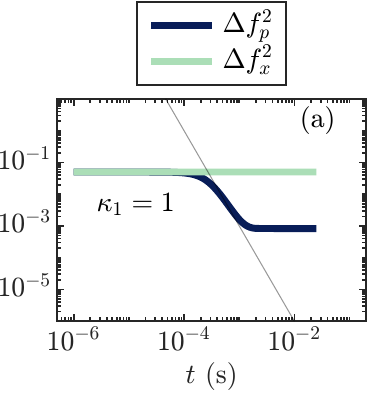}
    \includegraphics[trim={0cm \thebottrim cm 0cm 0cm}, clip]               {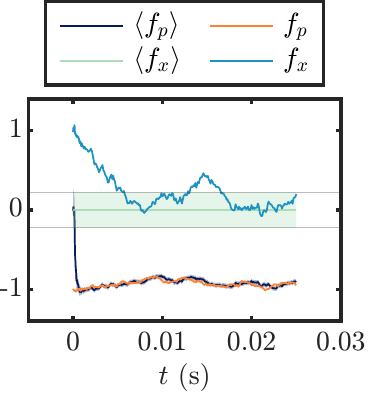} \\
    \includegraphics[trim={0cm \thebottrim cm 0cm \thetoptrim cm}, clip]    {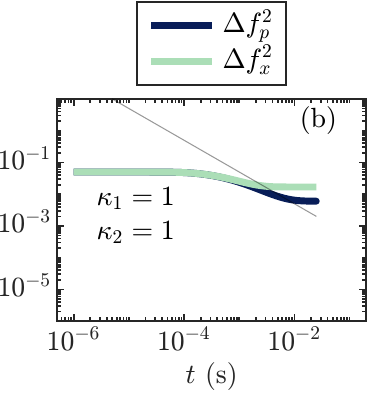}
    \includegraphics[trim={0cm \thebottrim cm 0cm \thetoptrim cm}, clip]    {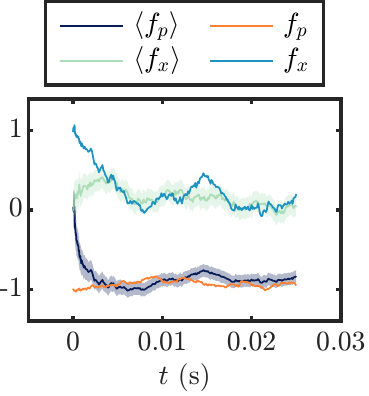} \\
    \includegraphics[trim={0cm \thebottrim cm 0cm \thetoptrim cm}, clip]    {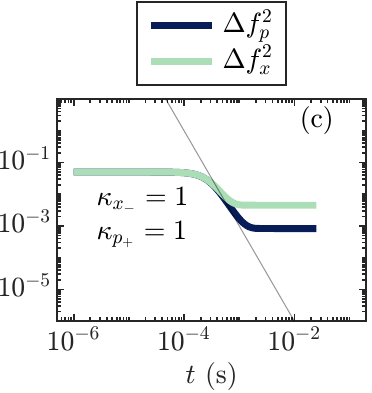}
    \includegraphics[trim={0cm \thebottrim cm 0cm \thetoptrim cm}, clip]    {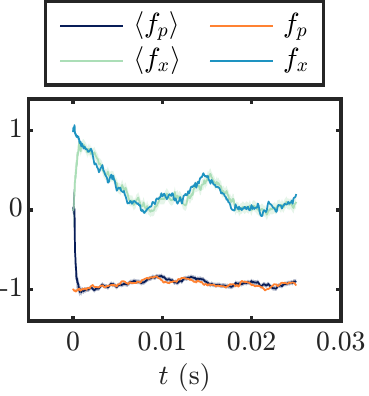} \\
    \includegraphics[trim={0cm  0cm 0cm \thetoptrim cm}, clip]              {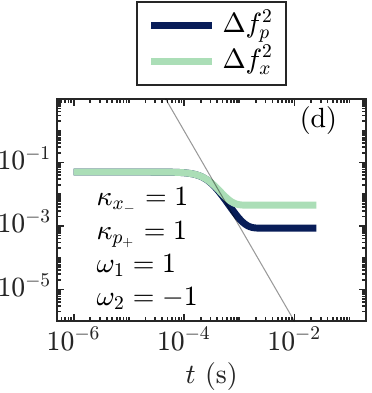}
    \includegraphics[trim={0cm  0cm 0cm \thetoptrim cm}, clip]              {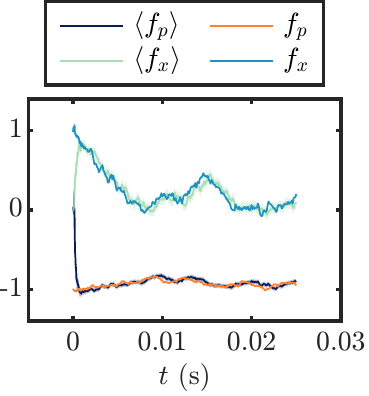} 
    \caption{
        Estimation of a fluctuating perturbations with $\gamma_{x,p}$ and $\sigma_{x,p}$ given in Table~\ref{tab:parametervalues}. 
        The left panels show the variance $\covar{}{f_i}$ with $1/t$ and  $1/t^3$ trend lines. The right panels show the true and estimated perturbation $\Braket{f_i} \pm  \std{}{f_i}$ (the true values are taken to be large positive and negative to distinguish their estimates clearly in the panels).  
        The upper panels (a) show the results for probing of $\p[S_1]$ for a single oscillator with $\omega_1=0$ being sensitive to only the value of $f_p$. 
        Panels (b) show the results for probing of $\x[S_1]$ and $\p[S_1]$ of  a single oscillator with $\omega_1=0$, being sensitive to both $f_x$ and $f_p$, but without the benefit of squeezing of the oscillator. 
        Panels (c) are for the probing and squeezing of EPR varables $\x[-]$ and $\p[+]$ of two oscillators with $\omega_1=\omega_2=0$, and hence the accelerated estimation of both $f_x$ and $f_p$. 
        The lower panels (d) are for the probing  of EPR varables $\x[-]$ and $\p[+]$ of two oscillators with opposite, non-vanishing frequencies. All listed parameters are given in units of the values in Table~\ref{tab:parametervalues}. The higher variance of $f_x$ is due to its faster OU damping and fluctuations.  
    }
    \label{fig:fluctfield}
\end{figure}

When the perturbations are fluctuating in a stochastic manner, early parts of long detection records do not contribute to the   estimator at later times and the variances approach final steady state values. 
The exact values and rates at which the steady state variances are approached depend on the measurement scenario as shown in Fig.~\ref{fig:fluctfield}. 

In Fig.~\ref{fig:fluctfield} we assume distinct true initial values for the OU processes, while the estimate assumes vanishing mean values and variances for  $f_x$ and $f_p$, equal to the OU steady state variances, $\frac{\sigma_x}{2\gamma_x} = \frac{\sigma_p}{2\gamma_p} = 0.05$ in our dimensionless unit. 

Figure~\ref{fig:fluctfield}(a) shows results for $\omega_1=0$ when we probe only the oscillator quadrature variable affected by $f_p$.  The variance of $f_x$ does not change, while the variance of the probed perturbation $f_p$ converges to a small steady state value. 
In Figure~\ref{fig:fluctfield}(b), we show results when both oscillator quadratures are probed and
$\omega_1=\omega_2=0$. Both perturbations are correctly estimated, but with larger variance than for $f_p$ in panel (a) due to the absence of squeezing. The range of values explored by $f_x$ and $f_p$ are the same, but the faster damping and diffusion of $f_x$ implies that it is more difficult to estimate precisely and hence it shows a larger variance. 

Probing both $\x[-]$ and $\p[+]$ for oscillators with $\omega_1=\omega_2=0$ leads to the results shown in panel (c), where both perturbations have variances that follow a cross over from an initial $1/t^3$ to a small constant variance like for $f_p$ in panel (a). The same behavior is seen in  Figure~\ref{fig:fluctfield}(d) for the probing of $\x[-]$ and $\p[+]$ for oscillators with opposite, finite frequencies $\omega_2=-\omega_1$. For the chosen parameters, we reach the constant steady state values before the cross over between the $1/t^3$and $1/t$ behavior observed in Fig.\ref{fig:cstfield}.

The successful demonstration of the efficient real time tracking of both fields using entangled oscillator variables with one positive and one negative evolution frequency constitutes a confirmation of the use of the back action evading protocol by Tsang and Caves \cite{tsang2010coherent,tsang2012evading} and is a main result of this paper. 

\subsection{Smoothing}

Finally, we apply the theory of past quantum states \eqref{eqs:PQS_moments} to retrodict the time dependent values of the perturbations based on the full measurements records. Such analysis was done for a single perturbation in 
Ref.~\cite{cheng2020estimating}, and  
it also applies to the negative mass oscillator setting and simultaneous estimation of two perturbations by two optical probes.

\begin{figure}[h]
\vspace{1cm}
    \centering
    \includegraphics[]{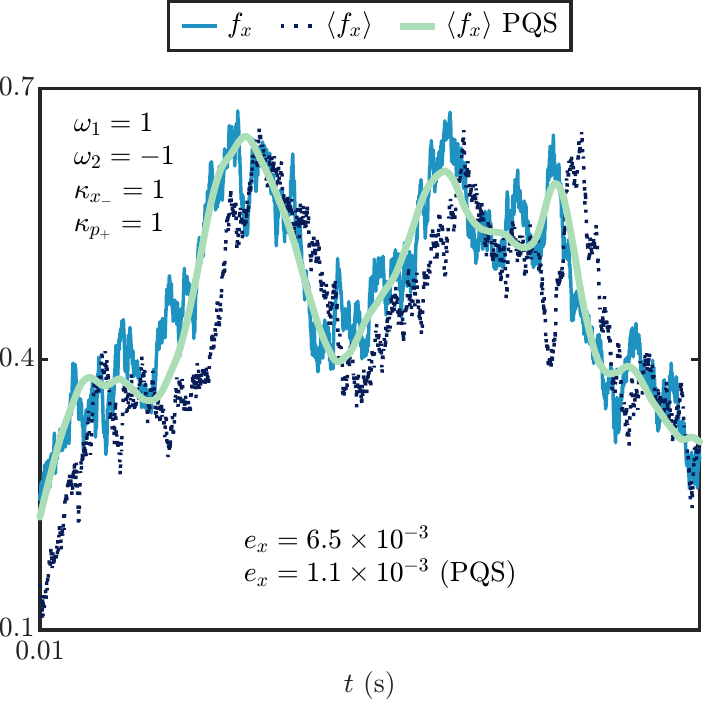}    
    \caption{Perturbation estimation using past quantum states.
    The mean-square error $e_x$ is recorded from $t_1 = 0.01\, \si{s}$ to $t_2 = T$. 
    If $t_1 = 0$ the error increases roughly by a factor of 2. 
    }
    \label{fig:pqs}   
\end{figure}

Figure~\ref{fig:pqs} 
compares the time dependent $f_x$ smoothed estimate with the true (simulated) time dependent value, shown as the lighter and darker (most noisy) solid curves. The forward filtering estimate is shown by the dotted data and follows the true time dependence but with a clearly visible delay, while the smoothed estimate (light solid curve) is visibly smoother than the forward filtering estimate, and it does not lag systematically behind the signal. 
The corresponding estimates match the true perturbation values better which can be quantified by the mean-square error $\overline{(f_i(t) - \Braket{f_i(t)})^2}$.
With the parameters chosen in Table~\ref{tab:parametervalues}, the smoothed estimate leads to a reduction of the mean square error by a factor $3-4$ compared to the forward filter estimate.  

\section{Conclusion}
In summary, we have presented a real time analysis of a back action evading measurement protocol based on EPR correlations between a probe oscillator and an ancillary negative mass oscillator \cite{tsang2012evading}. Continuous probing of the EPR variables enables their simultaneous squeezing and hence enhanced sensitivity to changes caused by perturbations of the position and momentum quadrature of the probe oscillator. The degree of squeezing of the quantum oscillators, the correlations between the oscillator variables and the perturbations, and the variance of our resulting estimate of the perturbations are all governed by a covariance matrix which obeys a deterministic equation, while their mean values, and hence the estimated value of the perturbation are continuously updated in accordance with the random measurement record. Finally, Bayesian estimates for the perturbations were obtained relying at every instant of time on both earlier and later measurement data. The procedure to describe these smoothed estimates yielded a further reduction in the uncertainty about the actual perturbation.   

Our theory builds on the conditional dynamics of a hybrid quantum classical density matrix subject to continuous measurements, and it is significantly simplified by the restriction to Gaussian states throughout the process. This restriction applies well to mechanical and field oscillators, but also to large polarized spin ensembles, and the assumption of interaction Hamiltonians with only quadratic terms applies exactly or to a good approximation for many studies with these systems.
We also note that systems with many degrees of freedom such as Bose-Einstein condensates and generic many-body systems may be well described by Bogoliubov theory or second order cumulant expansion methods and that our Gaussian covariance matrix method can readily deal with a large number of degrees of freedom and hence explore the prospects of sensing with such complex systems. There is, however, also a rich potential to explore non-Gaussian states for precision measurements \cite{evrard2019enhanced}, and, e.g., the assumption that the classical perturbations obey an Ornstein-Uhlenbeck process may be challenged in many practical sensing applications and hence their representation by Gaussian distributions become invalid. Integration of the hybrid quantum filtering with more pragmatic signal processing models, along the lines of Ref.~\cite{jimenez2018signal}, may then be needed for back action evading measurements to reach their full potential.

\section{Acknowledgements}
This work was supported by  the Danish National Research Foundation through the Center of Excellence for Complex Quantum Systems (Grant agreement No. DNRF156) and the  European  QuantERA grant C'MON-QSENS!, by Innovation Fund Denmark Grant No. 9085-00002.

\twocolumngrid
\bibliography{references}

\end{document}



\title{Supplementary material for \\ \textit{Gaussian theory for estimating fluctuating perturbations \\ with back action evasive oscillator variables}}

\author{Jesper Hasseriis Mohr Jensen}
 \email{jhasseriis@phys.au.dk}
\address{Department of Physics and Astronomy,
Aarhus University, Ny Munkegade 120, DK-8000 Aarhus C, Denmark} 
\author{Klaus M{\o}lmer}
\email{moelmer@phys.au.dk}
\address{Aarhus Institute of Advanced Studies, Aarhus University, H{\o}egh-Guldbergs
Gade 6B, DK-8000 Aarhus C, Denmark\\
Center for Complex Quantum Systems, Department of Physics and Astronomy,
Aarhus University, Ny Munkegade 120, DK-8000 Aarhus C, Denmark}


\date{\today}

\maketitle
\clearpage

This note presents supplementary derivations and results of the estimation procedures presented in Ref.~\cite{maintext}. Section \ref{sec:overview} elaborates on the hybrid quantum-classical description of the system. 
In Sec.~\ref{sec:bayesian} it is shown that the measurement back action in the augmented quantum state formalism is fully equivalent to Bayesian inference of the classical perturbations acting on the genuine quantum system. 
In Sec.~\ref{sec:wigner} we show that the hybrid Wigner function does not depend on ``dummy" quantum conjugate variables to the unknown classical perturbations 
(the Wigner function is independent of, and has infinite variance in those directions). This allows their exclusion from the formal description of both our forward filter theory and smoothed estimate based on the past quantum state moments. In Sec.~\ref{sec:analytical} we present some analytical results for the time evolution of quantum state and classical parameter variance.

\section{Hybrid quantum and classical Gaussian distributions}
\label{sec:overview}
The most general setup in the main text concerns two oscillators $S_1$ and $S_2$ with respective oscillation frequency $\omega_{1}$ and $\omega_2$ being continuously probed by two light meters $L_1$ and $L_2$ 
where the coupling between $S_i$ and $L_j$ is $\kappa_{ij}$ for $i,j=1,2$. 
The oscillator variables $\x[S_1]$ and $\p[S_1]$ are coupled to classical perturbations $f_x(t)$ and $f_p(t)$ with strength $c_i$ for $i=x,p$, while the perturbations evolve in time according to stochastic Ornstein-Uhlenbeck (OU) processes with diffusion $\sigma_i$ and decay rates $\gamma_i$. 

The relevant collective Einstein-Podolsky-Rosen (EPR) variables and their variances are related to the individual quantum oscillators $S_1$ and $S_2$ by
\begin{align}
\x[-] &= \frac{1}{\sqrt{2}}(\x[S_1] - \x[S_2]), && \covar{}{\x[-]} = \frac{1}{2}\left( \covar{}{\x[S_1]} + \covar{}{\x[S_2]} \right) -  \covar{\x[S_1]}{\x[S_2]}, \\
\p[+] &= \frac{1}{\sqrt{2}}(\p[S_1] + \p[S_2]), && \covar{}{\p[+]} = \frac{1}{2}\left( \covar{}{\p[S_1]} + \covar{}{\p[S_2]} \right) +  \covar{\p[S_1]}{\p[S_2]}.
\end{align}
These EPR variables commute, $[\x[-], \p[+]] = 0$, and both can be squeezed indefinitely which can be effected by the back action from the continuous coupling and measurement (successive homodyne detection of temporal segments  of the light field). 
Probing $\p[+]$ ($\x[-]$) is achieved by choosing $\kappa_{11} = \kappa_{21} \equiv \kappa_1$ ($\kappa_{12} = -\kappa_{22} \equiv \kappa_2$).
The adverse measurement back action effects from the anti-squeezed conjugate variables $\p[-], \x[+]$ is evaded by choosing $\omega_1 = -\omega_2$. 
We assume dimensionless oscillator variables $[\x[S_i], \p[S_j]] = i \delta_{i,j}$ and units where effectively $\hbar=1$. 

All the system components are Gaussian elements, i.e. the Wigner function is a multivariate Gaussian, and all the operations preserve the Gaussianity. 
This affords a compact description in terms of the mean vector $\vec a$ and covariance matrix $\mat A$ and their effective evolution (the sequentially incident light meter subsystems are traced out after each measurement).
That is, the elements of $\vec a$ and $\mat A$ are the quantum expectation values and covariances, respectively, of the oscillator variables at time $t$.

A hybrid classical-quantum formalism is employed to conveniently estimate the classical perturbations. 
This is achieved by associating the perturbation $f_i$ with eigenstates and  eigenvalues of an ancillary quantum operator $\op f_i$,  $\op f_i \ket{f_i} = f_i \ket{f_i}$. In our example we deal with two such perturbations, $i=x,p$. 
The new variables are introduced through an augmented density matrix, $\rhotop = \int \ket{f_x, f_p} \bra{f_x, f_p} \otimes \op \rho_{f_x,f_p}\; df_x df_p$, 
where each $\op \rho_{f_x,f_p}$ is an operator defined on the space of genuine quantum variables and has a time evolution governed by the application of the candidate values  of the perturbations and conditoned on the measurement outcomes of the light probes. 
The joint probability density for $f_x$ and $f_p$ being the true values of the perturbation is given by $P(f_x, f_p) = \tr[\op\rho_{f_x,f_p}]$. 
Since the total normalization is given by $\rhotop$, each measurement event redistributes the norm between the individual $\op \rho_{f_x,f_p}$'s, and hence the probabilities, in a Bayesian manner. 

To recover a Gaussian description we let the ancillary quantum operators 
be position operators $\op f_i = \x[f_i]$ with conjugate, uncoupled ``dummy" variables $\p[f_i]$.
Then there exists a Wigner function defined on the joint space of all the genuine and ancillary quantum variables. 
The OU process is described statistically by a Fokker-Planck equation and its diffusion and decay rates can be equivalently imposed on the associated ancillary quantum variables in a Gaussian manner.
This results in an augmented mean vector $\vec {\tilde{a}} = \vec {\tilde{a}}_\rho$ and covariance matrix $\mat{\tilde{A}} = \mat{\tilde{A}}_\rho$, which
include the first and second moments of our classical Gaussian estimator of the perturbations. 

The values of the perturbations  at time $t$ can also be retrodicted based on the entire detection record using the theory of past quantum states. 
This introduces a backward evolving effect matrix $\op{\tilde{E}}$ on the same Hilbert space as $\rhotop$ which also has a Gaussian Wigner representation with mean vector $\vec{\tilde{a}}_E$ and covariance matrix $\mat{\tilde{A}}_E$. 
Finally, appropriate combination of the $\rhotop$ and $\op{\tilde{E}}$ moments yields a vector $\vec{\tilde{m}}_{\rho, E}$ and a matrix $\mat{\tilde{A}}_{\rho, E}$. An element of $\vec{\tilde{m}}_{\rho, E}$ and the corresponding diagonal element of $\mat{\tilde{A}}_{\rho, E}$ yield the estimated mean and  variance of the outcome  of a projective measurement of the corresponding observable. These form our retrodicted Gaussian Bayesian estimator for the elements representing the classical perturbations.



\section{Augmented density matrix updates as Bayesian inference}
\label{sec:bayesian}
\newcommand{\phie}{\phi_\epsilon}
The classical-quantum hybrid formalism in the main text includes the perturbation $f$ to be estimated as an ancillary quantum operator $\op f$ on equal footing with the genuine quantum operators.  
This allows a convenient density matrix ansatz (we consider here a single perturbation and light meter for brevity), 
\begin{align}
    \op {\tilde{\rho}}  &= \int \ket{f}\bra{f} \otimes \op\rho_f \, df = \int \ket{f}\bra{f} \otimes \tr[\op\rho_f] \op\rho_f^{(n)} \, df, 
\end{align}
where the normalization of $\op\rho_f$ is explicitly factored, $\op\rho_f = \tr[\op\rho_f] \op\rho_f^{(n)}$. $\tr[\op\rho_f^{(n)}]=1$ and
$\tr[\op\rho_f]$ is the probability density for the perturbation to have the classical value $f$,
\begin{align}
    P(f) = \tr[\ket{f}\bra{f} \rhotop] =\tr\left[ \ket{f}\bra{f} \otimes \op\rho_f^{(n)}\right] \tr[\op\rho_f] = \tr\left[ \ket{f}\bra{f}\right] \tr[\op\rho_f^{(n)}] \tr[\op\rho_f] = \tr[\op\rho_f].
\end{align}
Note here that $\tr[\ket{f}\bra{f}] = \int \delta(f-f')^2\, df$ is technically divergent, in the same manner as the trace of the resulting resolution of the identity operator diverges due to the infinite dimensional Hilbert space. This issue may be formally dealt with by use of regular approximations to the delta functions and proper truncation of the Hilbert space, and the delta functions and the  
$\ket{f}\bra{f}$ do not pose problems when used in a consistent manner.

The unitary evolution of each $\op \rho_f$ is described by an operator $\op U_f$ where the perturbation takes the definite classical value $f$. 
The total unitary evolution operator for $\rhotop$ can then be constructed by
\begin{align}
    \op{\tilde U} &= \int \ket{f}\bra{f} \otimes \op U_f\,df.
\end{align}
A unitary evolution alone,
\begin{align}
\op{\tilde U} \rhotop \op{\tilde{U}}^\dagger = \int \ket{f}\bra{f} \otimes (\op U_f \op\rho\ut{(n)} \op U_f^\dagger)\, \tr[\op \rho_f]\, df,
\end{align}
does not change the probability distribution for $f$, 
\begin{align}
    P(f) = \tr[ \ket{f} \bra{f} (\op{\tilde U} \rhotop \op{\tilde U}^\dagger)] = \tr\Big[\ket{f}\bra{f} \otimes (\op U_f \op\rho \op U_f^\dagger)\ut{(n)}\, \tr[U_f \op \rho_f U_f^\dagger]\Big] = \tr[\op \rho_f].
\end{align}
Suppose now the unitary evolution is followed by a measurement, $\op \Omega_{q} = \op M_q \op{\tilde{U}}$, where $\op M_q$ is the POVM element for the outcome $q$, then the conditioned state is 
\begin{align}
    \rhotop_q &= \frac{\op M_q \op{\tilde{U}} \rhotop \op{\tilde{U}}^\dagger \op M_q^\dagger}{\tr[\op M_q \op{\tilde{U}} \rhotop \op{\tilde{U}}^\dagger \op M_q^\dagger]}.
\end{align}
The denominator is the probability for detecting the outcome $q$,
\begin{align}
    \tr[\op M_q \tilde U \rhotop \tilde U^\dagger \op M_q^\dagger] = \int \tr[\op M_q^\dagger \op M_q (\op U_f \op \rho_f \op U_f^\dagger)\ut{(n)}] \,\tr[\op \rho_f]\, df = \int P(q|f) P(f)\, df = P(q).
\end{align}
The conditioned state can therefore be written
\begin{align}
    \rhotop_q &= \int \ket{f}\bra{f} \otimes \left(\op M_q \op U_f \op \rho_{f}  \op U_f^\dagger \op M_q^\dagger \right) \frac{1}{P(q)}\, df
    = \int \ket{f}\bra{f} \otimes \op M_q (\op U_f \op \rho_{f}  \op U_f^\dagger)\ut{(n)}\op M_q^\dagger 
    \frac{\tr[\op\rho_f]}{P(q)}
    . 
\end{align}
Finally, the updated probability distribution for $f$ after detecting $q$ is
\begin{align}
    P(f|q) = \tr[\ket{f}\bra{f} \rhotop_q] = \tr\left[\op M_q^\dagger \op M_q (\op U_f \op \rho_{f}  \op U_f^\dagger)\ut{(n)}  \right] \frac{\tr[\op\rho_f]}{P(q)} = \frac{P(q|f) P(f)}{P(q)},
\end{align}
which is exactly Bayes rule. 
This result generalizes straightforwardly to an arbitrary number of perturbations and probe fields, and applies successively as the measurement record accumulates. 

\section{Augmented density matrix Wigner function}
\label{sec:wigner}
To admit the unknown classical perturbation into the quantum Wigner function description we specify $\op f = \x[f]$ as the position operator 
of an ancillary system with a conjugate momentum operator $\p[f]$.  
Then there exists a Wigner function $\W_{\tilde \rho}(x,p,x_f,p_f)$ (for simplicity of notation we consider a single perturbation and oscillator).

The ancillary system has a vanishing Hamiltonian and $\p[f]$ does not couple to $\x[f]$ or to the genuine quantum oscillators, and will hence play no role in the dynamics. Due to  $\op {\tilde{\rho}}$ being diagonal in the continuous $\x[f]$ eigenbasis, the Wigner  
function does not depend on the dummy variable, 
\begin{align}
\W_{\rhotop}&(x, p, x_f, p_f) \notag \\ 
    &= \frac{1}{(\pi \hbar)^2} \int \braket{x+q, x_f + q_f | \Big( \int dx_f' \ket{x_f'}\bra{x_f'} \otimes \op \rho_{f'} \Big) | x-q, x_f-q_f} e^{2ipq/\hbar}e^{2ip_fq_f/\hbar} dq dq_f \notag \\
    &= \frac{1}{(\pi \hbar)^2} \int \braket{x+q| \op\rho_{f'}|x-q} \braket{x_f + q_f | x_f''} \braket{x_f'|x_f-q_f} e^{2ipq/\hbar}e^{2ip_fq_f/\hbar} dq dq_f dx_f'  \notag \\
    &= \frac{1}{(\pi \hbar)^2} \int \braket{x+q| \op\rho_{f}|x-q} e^{2ipq/\hbar}dq \times  \delta\big((x_f-x_f') + q_f\big)\delta\big((x_f-x_f')-q_f\big) e^{2ip_fq_f/\hbar}    dq_f dx_f'. \label{eq:Wrhotilde}
\end{align}
The two delta-functions imply that the integral only acquires contributions for  $x_f=x_f'$ and $q_f=0$ and hence it is independent of $p_f$ and sifts out the dependence of $\rho_f$ on $f$.    
For Gaussian states this implies infinite variance in the $p_f$ direction, which however, plays no role in the dynamics. Being independent of $p_f$  also implies that we do not need to retain the argument in the Wigner function and in the mean values and covariances.
Generalizing to multiple perturbations and returning to the $\vec f$-notation, we hence
specify only the effective $\W_{\tilde \rho}(\vec{x,p,f})$ and associated joint Gaussian moments
$\mat{\tilde{a}}_{\rho}$ and $\mat{\tilde{A}}_{\rho}$ for the ancillary and genuine quantum oscillators. 



In the retrodictive past quantum state theory we had recourse to the Wigner function for the effect matrix, $\W_{\tilde E}(x,p,f,p_f)$, parametrized by a covariance matrix $\mat{\tilde{A}}_E$ and mean vector $\vec{\tilde{a}}_E$. To calculate the probability distribution $P\st{pqs}(f)$ conditioned on the entire detection record, it was necessary to determine Wigner functions for operators $\ket{f'} \bra{f'}\op{\tilde X}$ with $X=\rho, E$. Since $\op{\tilde \rho}$  and $\op{\tilde E}$ are both diagonal in the  $\ket{f}$ basis, the same holds for  
$\ket{f'} \bra{f'}\op{\tilde X}$, and their Wigner functions are both independent of $p_f$ and they evaluate readily to, 
%
\begin{align}
    \W_{\ket{f'}\bra{f'} \tilde\rho}(x,p,f) = \delta(f-f') \W_{\tilde{\rho}}(x, p, f'),
\end{align}
with a similar result for $\op{\tilde{E}}$ and straightforward generalization to more oscillators and perturbations.

\section{Numerical and analytical results}
\label{sec:analytical}
The evolution of the Gaussian covariance matrix and mean values due to the measurement of the quadrature variables of the probe field follows from the reduction of their joint quantum state by the measurement process. The Gaussian state formalism thus yields discrete analytical update formulae for the covariance matrix $\mat A$ and mean vector $\vec a$ (the ``$\sim$'' notation is omitted for brevity). 
The covariance matrix evolves independently of the actual measurement outcomes and in the limit of frequent probing, it solves a so-called Ricatti matrix differential equation,
\begin{align}
    \dot{\mat A} &= \lim_{\dt \rightarrow 0^+} \frac{\mat A\ut{}(t+\dt) - \mat A(t)}{\dt} \notag \\
    &\equiv \mat G - \mat D \mat A - \mat A\mat E - \mat A\mat F\mat A,
\end{align}
where, for our system, the matrices $\mat G, \mat D, \mat E, \mat F$ follow from the infinitesimal linear transformation of the Gaussian variables given in Ref.~\cite{maintext}, 
{
\footnotesize
\begin{alignat}{2}
\mat G &= 
\left(\begin{array}{cccccc}
{\kappa_{11} }^2  & 0 & \kappa_{11} \,\kappa_{21}  & 0 & 0 & 0\\
0 & {\kappa_{12} }^2  & 0 & \kappa_{12} \,\kappa_{22}  & 0 & 0\\
\kappa_{11} \,\kappa_{21}  & 0 & {\kappa_{21} }^2  & 0 & 0 & 0\\
0 & \kappa_{12} \,\kappa_{22}  & 0 & {\kappa_{22} }^2  & 0 & 0\\
0 & 0 & 0 & 0 & 2\,\sigma_x  & 0\\
0 & 0 & 0 & 0 & 0 & 2\,\sigma_p 
\end{array}\right), \quad 
&&\mat D = \left(\begin{array}{cccccc}
0 & -\omega_1  & 0 & 0 & -c_x  & 0\\
\omega_1  & 0 & 0 & 0 & 0 & -c_p \\
0 & 0 & 0 & -\omega_2  & 0 & 0\\
0 & 0 & \omega_2  & 0 & 0 & 0\\
0 & 0 & 0 & 0 & \gamma_x  & 0\\
0 & 0 & 0 & 0 & 0 & \gamma_p 
\end{array}\right), \\
\mat E &= \left(\begin{array}{cccccc}
0 & \omega_1  & 0 & 0 & 0 & 0\\
-\omega_1  & 0 & 0 & 0 & 0 & 0\\
0 & 0 & 0 & \omega_2  & 0 & 0\\
0 & 0 & -\omega_2  & 0 & 0 & 0\\
-c_x  & 0 & 0 & 0 & \gamma_x  & 0\\
0 & -c_p  & 0 & 0 & 0 & \gamma_p 
\end{array}\right), 
&&\mat F = \left(\begin{array}{cccccc}
{\kappa_{12} }^2  & 0 & \kappa_{12} \,\kappa_{22}  & 0 & 0 & 0\\
0 & {\kappa_{11} }^2  & 0 & \kappa_{11} \,\kappa_{21}  & 0 & 0\\
\kappa_{12} \,\kappa_{22}  & 0 & {\kappa_{22} }^2  & 0 & 0 & 0\\
0 & \kappa_{11} \,\kappa_{21}  & 0 & {\kappa_{21} }^2  & 0 & 0\\
0 & 0 & 0 & 0 & 0 & 0\\
0 & 0 & 0 & 0 & 0 & 0
\end{array}\right).
\end{alignat}}
\noindent
This is a nonlinear matrix equation for $\mat A$, but by writing $\mat A = \mat W \mat U^{-1}$ it can be decomposed into two linear matrix equations $\dot{\mat W} = -\mat D\mat W + \mat G\mat U$ and $\dot {\mat U} = \mat F \mat W + \mat E\mat U$. It is generally quite challenging to find closed-form solutions for $\mat A$, but we shall consider a few interesting limiting cases that illuminate our numerical findings.


\subsection{Squeezing by probing}
When there are no perturbations present, the last two rows and columns can be removed from the Ricatti matrices. 
We assume that the oscillator variables are initially uncorrelated with ground state variances, $[\mat A(0)]_{i,j} = \delta_{i,j}$, 
and we assume let $\omega_1 = \omega =  -\omega_2$, $\kappa_{11} = \kappa_{21} \equiv \kappa_1$, and $\kappa_{12} = -\kappa_{22} \equiv \kappa_2$.

Defining $K_\pm^2 = \kappa_2^2 \pm \kappa_1^2$, the time dependent EPR variances can be found and read
{
\small
\begin{align}
\covar{}{\x[-]}(t) &= \frac{
	\covar{}{\x[-]} + (
	\covar{}{\p[+]}- \covar{}{\x[-]})\sin^2\omega t + 2\covar{}{\x[-]}\covar{}{\p[+]} \left[K_+^2 t - \frac{\sin 2\omega t}{2 \omega}K_-^2\right]}
{
	\left[1+2\covar{}{\x[-]}K_+^2 t\right]
	\left[1+2\covar{}{\p[+]} K_+^2 t\right]
	+\frac{\sin \omega t}{\omega} K_-\left[\cos\omega t (\covar{}{\x[-]}- \covar{}{\p[+]}) - \frac{\sin\omega t}{\omega}K_- \covar{}{\x[-]}\covar{}{\p[+]}
	\right],
} \\
    \covar{}{\p[+]}(t) &= \frac{
	\covar{}{\p[+]} + (
	\covar{}{\x[-]}- \covar{}{\p[+]})\sin^2\omega t + 2\covar{}{\x[-]}\covar{}{\p[+]} \left[K_+^2 t + \frac{\sin 2\omega t}{2 \omega}K_-^2\right]}
{
	\left[1+2\covar{}{\x[-]}K_+^2 t\right]
	\left[1+2\covar{}{\p[+]} K_+^2 t\right]
	+\frac{\sin \omega t}{\omega} K_-\left[\cos\omega t (\covar{}{\x[-]}- \covar{}{\p[+]}) - \frac{\sin\omega t}{\omega}K_- \covar{}{\x[-]}\covar{}{\p[+]}
	\right],
}
\end{align}
}
where $\covar{}{\x[-]}$ and $\covar{}{\p[+]}$ on the right hand side are evaluated at $t=0$, and may be set to $1/2$ assuming initial  oscillator ground states.  For nonvanishing $\omega$, the asymptotic behaviour is the same for the two variables, $\covar{}{\x[-]}(t), \covar{}{\p[+]}(t) \propto (2 K_+^2 t)^{-1}$.
The equations apply for all values of $\omega$, while $\underset{\omega\rightarrow 0}{\lim}\sin(\omega t)/\omega {=} t$ must be applied and yields different asymptotic variances if different probing strengths are applied to the case of $\omega=0$.

\subsection{Estimation of a constant perturbation}

The case of separately probing two constant perturbations by the EPR variables is equivalent to the probing of a single constant perturbation and we get a similar result as in Ref.~\cite{molmer2004estimation} if we assume that the quantum oscillator variables have variances of $1/2$, $[\mat A(0)]_{i,j} =  \delta_{i,j}$, and that $\omega_1 = \omega_2=0$, while  
$\kappa_{11} = \kappa_{21} \equiv \kappa_1$ and $\kappa_{12} = -\kappa_{22} \equiv \kappa_2$. 
\begin{align} \label{eq:somklausoglars}
\covar{}{f_p}(t) = \frac{(1+2\kappa_{1}^2 t) \covar{}{f_p}}{1+2\kappa_1^2 t + \frac{2}{3}\kappa_1^2 c_p^2 \covar{}{f_p} t^3 + \frac{1}{3} \kappa_1^4 c_p^2  \covar{}{f_p} t^4}
\overset{t\rightarrow \infty}{\approx} \frac{6}{c^2 k_1^2} \times \frac{1}{t^3},
\end{align}
where $\covar{}{f_p}$ on the right hand side is evaluated at $t=0$, and an equivalent expression applies for $\covar{}{f_x}(t)$. 

\begin{figure}[H]
    \centering
    \includegraphics{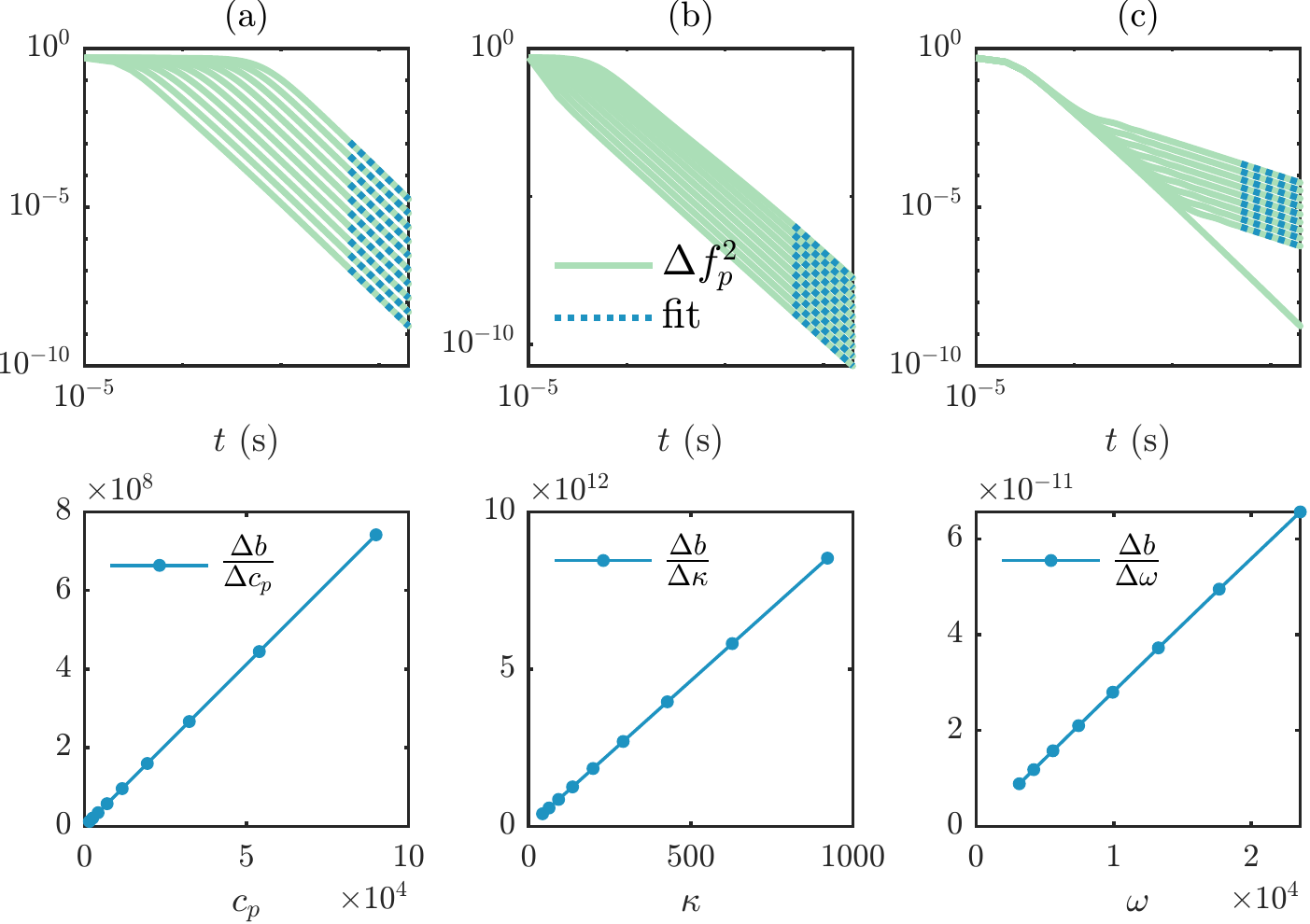}
    \caption{This figure shows the asymptotic behavior of $\covar{}{f_p}(t)$  in the upper parts of the panels. Each point in the lower panels is extracted from one of the curve fits for late times in the upper panels. The lower part of panel (a) shows, for $\omega = 0$, the variation with the perturbation parameter $c_p$ of the  fit of the asymptotic (rightmost) results to $1/(b t^3)$. The lower panel (b) shows the variation of the same fit as function of the probing strength parameter $\kappa$. Panel (c) shows the variation of the fit of the upper panel results with $b/t$ as function of $\omega$. 
    }
    \label{fig:fits}
\end{figure}

We are not able to solve the case of $\omega_1 = -\omega_2 \neq 0$ analytically. 
However, for large enough $t$ a crossover from $1/t^3$ to $1/t$  is observed  in our numerical calculations \cite{maintext}. 
This suggests the presence of an additional $\omega$-dependent $t^3$ term in the numerator of Eq.~\eqref{eq:somklausoglars}, and, by fitting the asymptotic behavior for different parameter values, Fig.~\ref{fig:fits} indicates that this $\omega$-dependence is quadratic, and for $\omega \neq 0$, $\covar{}{f_p}(t) \overset{t \rightarrow \infty}{\approx} \omega^2/(2 c_p^2 \kappa_1^2 t)$. 

\bibliography{supp_references}